\documentclass{aa} 
\usepackage{epsf} 
\usepackage{epsfig} 
\usepackage{color}

\newcommand{\mdot}{\mbox{$\dot{M}$}}

\newcommand{\vinf}{\mbox{$v_\infty$}}

\newcommand{\be}{\begin{equation}}
\newcommand{\ee}{\end{equation}}

\def\lesssim{\mathrel{\hbox{\rlap{\hbox{\lower4pt\hbox{$\sim$}}}\hbox{$<$}}}}
\def\gtrsim{\mathrel{\hbox{\rlap{\hbox{\lower4pt\hbox{$\sim$}}}\hbox{$>$}}}}
 
\begin{document}

\title{X-ray emission lines from inhomogeneous stellar winds}

\author{L.M. Oskinova, A. Feldmeier \and W.-R. Hamann}
\institute{Astrophysik, Universit\"at Potsdam, Am Neuen Palais 10, D-14469
Potsdam, Germany
}

\offprints{lida@astro.physik.uni-potsdam.de}
\date{Received date; Accepted date}
\authorrunning{Oskinova, Feldmeier, Hamann}
\titlerunning{X-rays from inhomogeneous winds}

\abstract{It is commonly adopted that X-rays from O stars are produced
deep inside the stellar wind, and transported outwards through the bulk
of the expanding matter which attenuates the radiation and affects the
shape of emission line profiles. The ability of the X-ray observatories
Chandra and XMM-Newton to resolve these lines spectroscopically provided
a stringent test for the theory of the X-ray production. It turned out
that none of the existing models was able to fit the observations
consistently. The possible caveat of these models was the underlying
assumption of a smooth stellar wind. Motivated by the various evidences
that the stellar winds are in fact structured, we present a 2-D
numerical model of a stochastic, inhomogeneous wind. Small parcels of
hot, X-ray emitting gas are permeated with cool, absorbing wind material
which is compressed into thin shell fragments. Wind fragmentation alters
the radiative transfer drastically, compared to homogeneous models of
the same mass-loss rate. X-rays produced deep inside the wind, which
would be totally absorbed in a homogeneous flow, can effectively escape
from a fragmented wind. The wind absorption becomes wavelength
independent if the individual fragments are optically thick. The X-ray
line profiles are flat-topped in the blue part and decline steeply in
the red part for the winds with short acceleration zone. For the winds
where the acceleration extends over significant distances, the lines can
appear nearly symmetric and only slightly blueshifted, in contrast to
the skewed, triangular line profiles typically obtained from homogeneous
wind models of high optical depth. We show that profiles from a
fragmented wind model can reproduce the observed line profiles from 
$\zeta$\,Orionis. The present numerical modeling confirms the results
from a previous study, where we derived analytical formulae from a
statistical treatment.
}

\maketitle

\section{Introduction}

Hot massive stars possess strong stellar winds, as discovered by the
advent of UV spectroscopy (Morton \cite{Morton67}). One decade later
it was detected that these stars emit X-rays (Harnden et al.\ \cite{Harnden79},
Seward et al.\ \cite{Seward79}). The origin of these X-rays is debated
controversially. One of the first suggestions was the existence of a
hot corona close to the stellar photosphere. However, the surrounding
stellar wind should imprint strong K-shell absorption edges on the
X-ray spectrum, which were not observed (Cassinelli \& Swank
\cite{CS83}). The base-coronal model was finally ruled out because of
the lack of characteristic coronal emission lines (Baade \& Lucy
\cite{Lucy82}). Therefore it became commonly adopted that X-rays from
hot stars are produced within their stellar winds, although most of
the wind gas is obviously ``cool''.

The driving mechanism for the mass-loss from OB stars has been
identified with radiation pressure on spectral lines. A corresponding
theory was developed by Castor et al.\ (\cite{CAK75}, ``CAK''). After
a number of improvements, this theory seems to predict the observed
mass-loss rates and wind velocities correctly 
(Pauldrach et al.\ \cite{Pauldrach86}). However, it was pointed out 
early (Lucy \& Solomon
\cite{LucySolomon70}), and later further investigated (Carlberg
\cite{Carlberg80}; Owocki \& Rybicki \cite{OR84}), that the stationary
solution for a line-driven wind is unstable; small perturbation should
grow quickly and form shocks. This ``de-shadowing instability'' is
implicitly suppressed in the CAK theory.

The most detailed hydrodynamic modeling of this line driven
instability was presented by Feldmeier et al.\ (\cite{AF97}). These
calculations show how initially small perturbations grow and form
strong shocks that emit X-rays. This ``wind shock model'' was
principally able to reproduce the X-ray flux from the O supergiant
$\zeta$~Ori\,A, which had been observed at that time only at low
spectral resolution.

The ultimate test for models of X-ray production in stellar winds
became possible with the launch of the X-ray observatories Chandra and
XMM-Newton. Their spectrographs allow for the first time to resolve
X-ray emission line profiles. Surprisingly, the shape of the X-ray
line profiles turned out to be strikingly different among the few
individual O stars observed so far:

\begin{description}
\item
{$\theta^1$~Ori\,C (Schulz et al.\ \cite{Schulz00}) and $\zeta$~Ori\,A
(Waldron \& Cassinelli \cite{WaldCass01})} show profiles which appear
wind-broadened, symmetric and unshifted with respect to the center
wavelength;

\item
{$\delta$~Ori\,A (Miller et al.\ \cite{Milleretal02})} shows profiles
which are narrower, symmetric, and also centered at the laboratory
wavelength;

\item
{$\zeta$~Pup (Cassinelli et al.\ \cite{Cass01}; Kahn et al.\
\cite{Kahnetal01})} shows strongly broadened profiles which are
blueshifted against the rest wavelength but otherwise appear to be
symmetric.

\end{description}

If X-rays are produced deep inside the stellar wind, they have to
propagate through the absorbing cool wind before they can emerge
towards the observer. The opacity at X-rays energies is much larger
than in the visual and UV, due to the K-shell absorption of abundant
metal ions and the He\,{\sc ii} edge.  Red-shifted photons of an
emission line originate in the back hemisphere of the wind, and suffer
much stronger absorption than photons in the blueshifted part of the
line profile. Hence, as soon as the absorbing wind has a significant
optical depth, the line profiles should be blueshifted and skewed,
which means their shape resembles that of a triangle, with the flux
maximum at maximum blueshift (MacFarlane et al.\
\cite{MacFarlane91}). None of the observed stars meets this
prediction.

Kramer et al.\ (\cite{KCO03}) were able to fit eight X-ray lines
observed in $\zeta$\,Puppis with a homogeneous wind model. However,
these fits could only be achieved by assuming much less absorption in
the wind than expected from the generally adopted mass-loss rate of
this star.  Moreover, Kramer et al.\ (\cite{KCO03}) state that there
is not a strong trend of absorption with wavelength, in contrast to what
is expected from the energy dependence of the mass absorption
coefficient. Thus even for the prototype single O star $\zeta$~Puppis
the wind shock model fails to reproduce observed X-ray emission lines
consistently. At this point the question arose whether this model is
adequate.

We emphasize that the whole interpretation of X-ray spectra had been
based so far on over-simplified models, assuming a {\em homogeneous}
distribution of the absorbing material. By contrast, there is strong
empirical evidence that stellar winds are in fact strongly clumped
(e.g.\ Hamann \& Koesterke \cite{HK98}, Eversberg et
al.\ \cite{Ev98}, Puls et al.\ \cite{Puls03}).  This supports
hydrodynamic models (Owocki et al.\ \cite{OCR88}, Feldmeier
\cite{AF95}) which show that most of the material in a radiatively
driven wind is compressed into a series of dense shells by the
instability. As the hydrodynamic models are one-dimensional, they
cannot tell anything about the lateral structure of these wind
inhomogeneities. The de-shadowing instability acts only in the radial
direction; on the other hand, there is no obvious mechanism which
could synchronize the wind in different directions. The observed
variability in wind lines is not very pronounced, suggesting that the
dense shells break up into a large number of fragments.

Wind inhomogeneity alters the radiative transfer significantly. We
have studied the effects of wind fragmentation on the X-ray line
formation in Feldmeier et al.\ (\cite{PaperI}, Paper\,I), using a
statistical approach that holds in the limit of infinitely many
fragments. We showed analytically that the continuum wind opacity is
greatly reduced in a structured wind, compared to a smooth wind with
the same mass-loss rate. The line profiles we obtained are broad,
blueshifted and flat-topped, i.e. symmetric, and thus of promising
similarity to those observed in $\zeta$~Pup.

In the present paper we construct a stochastic wind model using a
numerical approach. This allows to drop a number of idealizations in
favor of a more realistic description.

\begin{description} 
\item {\it Statistical approach.} While Paper\,I relied on the
statistical limit of very many fragments, we allow now for an
arbitrary scale of fragmentation.

\item {\it Wind geometry.} We study models with different spatial
distributions of fragments, both in the  radial and lateral direction.

\item {\it Wind velocity law.} We abandon the restriction to a
constant wind velocity and allow for so-called $\beta$-velocity laws.
\end{description}

Last not least, another purpose of the present paper is to better
understand some surprising results obtained in Paper\,I. Despite of
our focus on X-ray emission line formation, we want to emphasize their
general importance for radiative transfer in inhomogeneous media.

In the next section we will introduce the stochastic wind model.
Section\,3 provides the formalism for solving the radiative transfer
in a fragmented wind, as implemented in our numerical modeling. The
basic effects of structured absorption on line profiles are
illustrated in Sect.\,4 using simplified examples. In Sect.\,5 we
introduce the concept of an effective opacity, and compare the line
profiles obtained numerically from the stochastic wind model with the
results from an analytical, statistical treatment. In Sect.\,6 we
evaluate line profiles using realistic assumptions and explore the
parameter space. Conclusions are drawn in the final Sect.\,7. A
forthcoming paper will be devoted to detailed fits of observations.

\section{The model}
\label{sect:the_model}

\subsection{Basic concepts}

Our model refers to the structure and physical conditions in the wind
shock model according to the hydrodynamic simulations by Feldmeier et
al.\ (\cite{AF97}). These dense flows driven by radiation pressure on
spectral lines are highly supersonic. The de-shadowing instability
leads to strong gas compression. Feldmeier et al.\ (\cite{AF97})
considered photospheric turbulence as seed perturbation of unstable
growth. The compressed regions appear as thin shells, moving outward
with nearly the stationary wind speed. Due to the lack of
synchronization between different radial directions, the shells
probably break up into fragments of small lateral extent. The space
between fragments is essentially void, but at the outer side of the
dense shells exist extended gas reservoirs. From their outer edge,
small gas cloudlets are ablated and accelerated to high speed by
radiation pressure. Propagating through an almost perfect vacuum, they
catch up with the next outer shell and ram into it. In this collision,
the gas parcels are heated and emit thermal X-radiation. This
mechanism works within a limited radial range of the wind acceleration
zone.

The two structural wind components of highly compressed, cool
fragments and hot X-ray emitting gas parcels are disjunctive. X-ray
emission ceases when the wind reaches its terminal speed. On the other
hand, despite the expansion due to the internal pressure, the cool
fragments are maintained to distances of a few tens of the stellar
radius, before they gradually dissolve into a homogeneous outflow.

\subsection{Assumptions}

Our stochastic model for X-ray line formation is designed to describe
in a generic way the structures and physical conditions of the stellar
wind, as summarized in the preceding subsection:

\begin{enumerate}

\item The flow is spherically symmetric and stationary in a statistical
sense. The matter propagates according to a monotonic velocity law.

\item The initially homogeneous wind becomes fragmented at some
specified distance from the stellar photosphere. The fragmentation is
maintained out to a distance $r_{\rm sh}^{\rm max}$ at which the
fragments dissolve into a homogeneous outflow. In the radial interval
$(r_{\rm sh}^{\rm min}, r_{\rm sh}^{\rm max})$ all wind material is
condensed into a finite number of discrete fragments of spherical
shells which propagate in radial direction.

\item All shell fragments cover the same solid angle as seen from the
center of the star, and preserve this angle during their
propagation. In contrast to Paper\,I this solid angle is not
assumed to be infinitesimally small.

\item We consider different statistical distributions of the cool
fragments: ``concentric spheres'', ``broken spheres'', ``random
fragments'', and ``cones''. These models are described in Sect.\,2.3.
The statistical approach of Paper\,I corresponds to the ``random
fragments'' distribution.

\item The X-ray emission originates from parcels of hot gas which are
randomly distributed. Their location is confined to the radial range
$r_{\rm em}^{\rm min}, r_{\rm em}^{\rm max}$. There is no
self-absorption by emitting material and no re-emission of X-rays
which became absorbed by cool fragments. Hence, emission is decoupled
from absorption, which allows for a formal solution of the radiative
transfer equation.

\end{enumerate}

Besides these physical assumptions, we restrict the model to
two-dimensional symmetry. Throughout the paper the index $j$ is
used when addressing cool absorbing fragments, while the index $i$ refers to 
parcels of X-ray emitting hot gas.

\subsection{Distribution of absorbers}
\label{sect:distribution_of_absorbers}

We introduce cartesian coordinates as follows. The $z$ axis points
towards the observer, while the impact parameter $p$ stands
perpendicular to the latter. Alternatively, a point $(p,z)$ is specified
by spherical coordinates, i.e.\ the radius $r=\sqrt{p^2+z^2}$ and the angle
$\vartheta=\cos^{-1}\mu$ between the line of sight and the radial
vector. Hence $\mu = z/r$. 

For the velocity we adopt the standard ''$\beta$-law''
\begin{equation} 
v(r)= v_\infty\ \left(1-\frac{r_0}{r}\right)^\beta,
\label{eq:vr} 
\end{equation} 
where $v_\infty$ is the terminal velocity, and $r_0$ is chosen such
that $v(r\!=\!R_\ast)=0.01\,v_\infty$. $R_\ast$ is the stellar
(photospheric) radius. For a given mass loss rate \mdot, the
continuity equation defines the density stratification via
$\rho(r)=\mdot/(4\pi r^2 v(r))$. We consider {\em continuum}
absorption of X-rays by the cool wind component, caused predominantly
by bound-free and K-shell ionisation processes. The opacity scales
linearly with the absorber density. Assuming for simplicity that the
relative population numbers of the absorbers do not vary with radius,
the opacity $\chi$ scales with the mass density $\rho(r)$,
\begin{equation} 
\chi(r) = \kappa\,\rho(r)
\label{eq:kappa} 
\end{equation} 
where $\kappa$ is the mass absorption coefficient which can be
calculated from the chemical composition, population numbers and the
continuum cross sections at the frequency of the considered X-ray
line.  To avoid explicit reference to these quantities, we specify the
wind by its total radial optical depth,
\begin{equation} 
\tau_\ast =
\int_{R_*}^{\infty} \chi(r)\ {\rm d}r\ . 
\label{eq:tast} 
\end{equation}

For a homogeneous wind, the optical depth $\tau_{\rm h}(p,z)$ 
between a point (p,z) and the observer is
\begin{equation}
\tau_{\rm h}(p,z) = \frac{\kappa \mdot}{4\pi}
\int_{z}^{\infty}\frac{{\rm d}z'}{{r'^2 v(r')}}\ .
\label{eq:tauhz}
\end{equation}
with $r' = \sqrt{p^2 + z'^2}$.

\begin{figure}[hbtp] 
\epsfxsize=\columnwidth
\centering \mbox{\epsffile{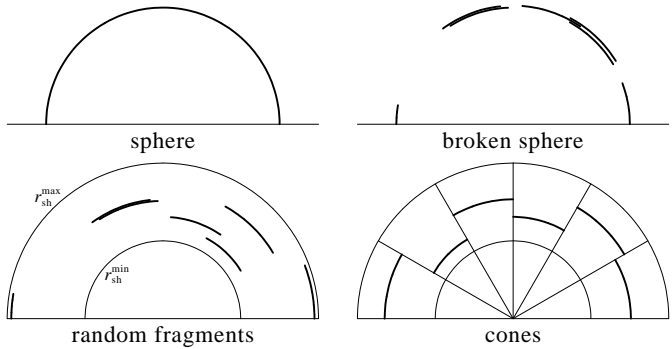}} 
\caption{Sketch of the distribution of absorbing fragments for
different models, for $\langle N_r\rangle =1$.}
\label{fig:modsketch}
\end{figure}

Next we proceed to the fragmented wind. We introduce different
versions of our model, according to different spatial arrangements of
the compressed fragments (see Fig.\,\ref{fig:modsketch}).

\medskip
\noindent{\bf Model ''concentric spheres''.}~First, we consider a
model where the matter is radially compressed into $N_r$ discrete
shells forming full concentric spheres.  In order that the
time-averaged mass-flux \mdot\, of shells resembles that of a
stationary, homogeneous wind, the probability to find a shell in the
radius interval $r, r\!+\!{\rm d}r$ must scale with $1/v(r)$. A
corresponding set of $N_r + 1$ random radii is generated by means of
the {\em von Neumann rejection method} (e.g.\ Press et al.\
\cite{NumRec92}). All the matter located between two subsequent radii
is compressed into a thin shell, moving with the stationary wind
speed. The mass in each shell is conserved with time.

The total mass of a homogeneous wind enclosed between two subsequent 
random radii $r_{\rm a}$ and $r_{\rm b}$ is
\begin{equation} 
M_j = \int_{r_a}^{r_b} 
       4\pi r^2 \rho(r)\ {\rm d}r \ =\  
      \mdot \int_{r_a}^{r_b}\frac {{\rm d}r}{v(r)}\ .  
\end{equation}
In the fragmented wind model, this material is swept up into a thin shell
of infinitesimal thickness $\Delta r$ located at $r_j$. The volume of
this shell is $V_j = 4\pi\ r_j^2\ \Delta r$, its matter density is
$M_j/V_j$, therefore the radial optical depth across the shell becomes
\begin{equation}
\label{eq:taufrag}
\tau_j^{\rm rad} = \frac {\kappa M_j}{4\pi r_j^2} = 
      \frac {\kappa \dot{M}}{4\pi r_j^2} 
      \int_{r_a}^{r_b}
      \frac{{\rm d}r}{v(r)}\ .
\end{equation}
Note that the shell thickness $\Delta r$ serves only as an auxiliary
variable, and cancels in the expression for $\tau_j$. Finally, the
location of the shell, $r_j\in\,[r_a, r_b]$, is determined by the
condition that this radial optical depth should be the same as for the
homogeneous wind when considering the radial range from where the
material has been swept up:\
\begin{equation}
\label{eq:tauhom}
\tau_{\rm h}^{\rm rad} = 
      \int_{r_a}^{r_b}
      \chi(r) {\rm d}r\ =\
      \frac {\kappa \dot{M}}{4\pi} 
      \int_{r_a}^{r_b}
      \frac{{\rm d}r}{r^2 v(r)}\ .
\end{equation}
Requiring $\tau_j^{\rm rad}=\tau_{\rm h}^{\rm rad}$ from
Eqs.\,(\ref{eq:taufrag}) and (\ref{eq:tauhom}), the shell location is
given by
\begin{equation}
r_j^2 = \int_{r_a}^{r_b} \frac{{\rm d}r}{v(r)}
        \left/
        \int_{r_a}^{r_b}\frac{{\rm d}r}{r^2 v(r)}
        \right. 
\label{eq:dsh}
\end{equation} 
In case of constant velocity, the shell location is just the harmonic
mean between the boundaries, $r_j=\sqrt{r_a \,r_b}$.

Thus, if shells are distributed according to Eq.\,(\ref{eq:dsh}),
where $r_a$ and $r_b$ are two random numbers, not only the mass loss
rate is the same as for the homogeneous wind but so is the radial
optical depth.
 
\medskip
\noindent{\bf Model ``broken spheres''.}~As a next step, the
absorbing spheres from the previous section are cut into pieces of
equal size.  In the actual 2-D cross section we consider, we cut each
ring into $N_\vartheta$ arcs of angle $2\pi/N_\vartheta$. Next, the
angular positions $\vartheta_j$ of the arcs are randomly distributed
between $0$ and $2\pi$ with uniform probability. Note that an
independent set of angles $\vartheta_j$ is generated for each
previously diced $r_j$. The fragments partially overlap, leaving some
radial directions free.

\medskip
\noindent{\bf Model ``random fragments''.}~In this model both
coordinates $r_j$ and $\vartheta_j$ are chosen randomly and
independent from each other. The probability distribution is
uniform in radius and in angle. In the statistical limit $N_\vartheta
\rightarrow \infty$, and assuming constant wind speed, this model was
studied analytically in Paper\,I.

\medskip
\noindent{\bf Model ``cones''.}~One might doubt whether in a real
stellar wind the radial and angular distributions of fragments can be
statistically independent, as assumed in the random fragment
model. One expects motions to be radial with negligible transversal
component. The inner, not yet fragmented zone of the wind provides a
steady, homogeneous, and angle-independent mass flux. The random
fragment model conserves this mass only in the statistical sense, as
at any moment the number of fragments in a given radial direction may
differ from that in another direction. In a snapshot of the random
fragments model, the radial optical depth is therefore different in
different directions. In order to investigate whether this has any
puzzling consequences, we put forward an alternative model where the
radial optical depth is strictly the same in all directions, as in the
concentric spheres model. To achieve this, we divide the angle
interval $[0, 2\pi]$ into $N_\vartheta$ equal segments.  Then sets of
$N_r+1$ random radii $r_j$ are generated independently for each of
these ``cones'' according to the description given for the
``concentric spheres'', and the fragments are placed accordingly. We
emphasize that each fragment is confined exactly to its
cone. Therefore the radial optical depth and the mass in each radial
direction is identical to that of a homogeneous wind at any instant of
time.  

\subsection{Distribution of emission}

The emitting gas is also distributed stochastically. We assume that a
large but finite number, $N_{\rm em}$, of small parcels of emitting
gas are randomly distributed in the stellar wind in the radial
interval ($r^{\rm min}_{\rm em}$, $r^{\rm max}_{\rm em}$). It is assumed for
simplicity that the emitting zones maintain their identity during
propagation with the velocity law $v(r)$; this implies that the
probability to find an emitting zone in the radial interval
($r,\,r\!+\!{\rm d}r)$ scales with $v^{-1}$. We assume the same velocity
law for both emission and absorption, and determine the location of
emitting zones again by von Neumann's rejection method.

\begin{figure}[hbtp] 
\epsfxsize=\columnwidth
\centering \mbox{\epsffile{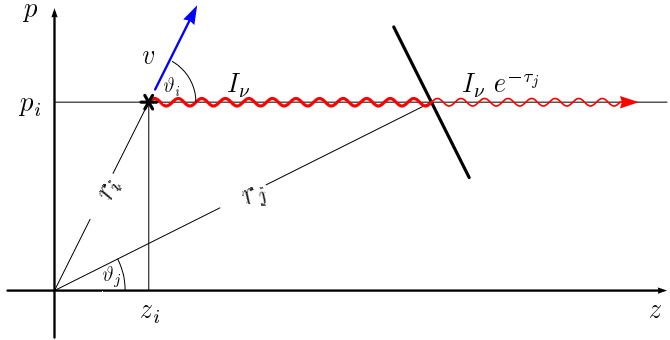}} 
\caption{Sketch of the coordinate system for the fragmented wind
model. X-ray line photons are emitted by hot, small gas parcels. Shown
is the emission site $i$ located at random coordinates $(p_i, z_i)$,
and moving with velocity $v(r)$ in the radial direction. The cool wind
consists of compressed and aligned fragments that are also randomly
distributed. The radiation from parcel $i$ towards the observer at $z
\rightarrow \infty$ has to cross a shell fragment $j$ at radius $r_j$
and covering an angle interval $[\vartheta_j- \frac{1}{2}\Delta
\vartheta, \vartheta_j+ \frac{1}{2}\Delta\vartheta]$. The optical
depth across fragment $j$ for the ray in $z$ direction is $\tau_j$,
reducing the line intensity $I_\nu$ by a factor $\exp (-\tau_j)$.}
\label{fig:coord}
\end{figure}

After the radius $r_i$ of an emitting parcel is selected, the random
choice for $\vartheta_i$ is to be made. For a uniform distribution
over the sphere, the probability distribution of $\vartheta$ must
scale with $\sin \vartheta$. This corresponds to a uniform probability
distribution of $z$ in the interval $(-r_i, r_i)$. As our model is
rotationally symmetric around the $z$ axis, we put the azimuth to
$\varphi_i = 0$.  Figure\,\ref{fig:coord} shows one of these
line-emitting zones, identified by its index $i$, and located at
spherical coordinates $(r_i,\vartheta_i)$, corresponding to Cartesian
coordinates ($z_i, p_i$). Actually $\vartheta_i$ does not appear
explicitly in our formalism, but only $r_i$, $z_i$ and $p_i =
\sqrt{r_i^2 - z_i^2}$ do.

A particular X-ray emitting zone with index $i$ has a line luminosity
$L_i$. We make the following simplifying assumptions. All emitting
parcels of gas contain the same amount of matter at the same
temperature. The line emission is powered by collisional excitation
and therefore scales with the density-squared. During motion, each
zone expands according to the continuity equation. Hence, the emission
$L_i$ of zone $i$ at radius $r_i$ scales with density,
\begin{equation}
\label{eq:L_i}
L_i = \frac{L_0}{r^2_i v(r_i)},
\end{equation}
with a constant $L_0$ determined by the line emissivity and the filling
factor. 

The total X-ray line luminosity internally released by the wind is
\begin{equation}
\label{eq:L_X}
L_{\rm X,int} = \sum_{i=1}^{i_{\rm max}} L_i\ .
\end{equation}
Because of absorption by shell fragments, only a fraction of this
radiation will emerge from the atmosphere.

\section{Radiative transfer}

After the X-ray line emitting zones and the continuum absorbing shell
fragments have been specified, the radiative transfer is calculated in
order to obtain the emergent X-ray line profile. In our model emission
is decoupled from absorption, which allows for the formal solution of
the radiative transfer equation. Formulating the radiative transfer in
the usual notation (cf.\ Mihalas \cite{Mihalas78}), we define the
emissivity $\eta_\nu$ of the X-ray emitting zone with index $i$ as
\begin{eqnarray}
\label{eq:emissivity}
\lefteqn{\eta_\nu (\nu_{\rm cmf},p,z,\varphi) =} \\ \nonumber
&& \frac{L_i}{4\pi}\ \delta(p-p_i)\ \delta (z-z_i)\ 
\delta(p_i(\varphi-\varphi_i))\
\delta(\nu_{\rm cmf} - \nu_0)\ .
\end{eqnarray}
In order to take advantage of the rotational symmetry of our model
around the $z$ axis, cylindrical coordinates $(p,z,\varphi)$ are used
here. The delta functions account for the negligible spatial extension
of the zone and the negligible intrinsic line width.  All delta
functions are normalized with respect to the integral over their
argument, $\int \delta(p-p_i) {\rm d}p = 1$, etc. Note that the
radiation is emitted at the line center frequency $\nu_0$ as measured
in the frame of reference co-moving with the flow. The corresponding
frequency $\nu$ as seen by an observer at $z \rightarrow \infty$ is
Doppler-shifted,
\begin{equation}
\nu_{i,{\rm obs}} = 
\nu_0 \left(1 + \frac{z_i}{r_i} \frac{v(r_i)}{c} \right)\ .
\end{equation}
If there were no absorption, the observer would receive an intensity
\begin{equation}
\label{eq:I-em}
I_{\nu,i} (\nu,p,\varphi) = \frac{L_i}{4\pi}\ 
\delta(p-p_i)\ \delta(p_i(\varphi-\varphi_i))\ \delta(\nu - \nu_{i,{\rm obs}})
\end{equation}
as obtained by integrating Eq.\,(\ref{eq:emissivity}) over $z$. 

It may happen that X-rays emitted from the considered parcel of hot
gas have to cross one ore more absorbing shell fragments while
traveling in the direction towards the observer. Thus we have first to
find out which fragments cover the impact parameter $p_i$ of the
emission spot. Imagine, the shell fragment with index $j$ meets this
condition (cf.\ Fig.\,\ref{fig:coord}). The radial optical depth
assigned to this fragment is $\tau^{\rm rad}_j$.  If the center of
this fragment has spherical coordinates $(r_j, \vartheta_j)$, the
optical depth actually experienced by the X-ray is
\begin{equation}
\tau_j = \frac{\tau^{\rm rad}_j}{|\mu_j|}\ .
\label{eq:tsh}
\end{equation}
This accounts for the longer geometrical path of rays which intersect
the shell fragment at an inclined angle. For simplicity we apply a
constant optical depth $\tau_j$ over the whole shell fragment,
neglecting its curvature.

Let us compile a list ${\mathcal J}_i$ which contains the indices $j$
of those shells which are hit by the ray from emission zone $i$. Then
the total optical depth for the ray from this zone, $\tau_i$, is the
sum over all absorption shells crossed, i.e.\
\begin{equation}
\tau_i = \sum_{j \in {\mathcal J}_i} \tau_j\ .
\label{eq:nh}
\end{equation}

The crucial point to understand is that the optical depth $\tau_i$
differs essentially from the optical depth of an individual fragment
$\tau_j$. The optical depth across a fragment depends on the
opacity $\chi(r)$ and the amount of swept-up material. In contrast,
the optical depth along a specific line of sight, $\tau_i$, depends in
the first place on the geometrical distribution of absorbing
fragments, which defines the list ${\mathcal J}_i$ over which the sum
in Eq.\,(\ref{eq:nh}) is taken. If, for example, the fragments are
located in such a way that the ray from a particular emission zone $i$
towards the observer does not cross any fragment, then $\tau_i=0$ for
this zone, even if all fragments have $\tau_j\gg 1$. Vice versa, if
there are many fragments of small $\tau_j\ll 1$ on the line of sight,
the summed-up optical depth for that ray, $\tau_i$, can be
significant.

For completeness, we have to specify the absorption of those parts of
the wind which are not considered as being fragmented. First, the
stellar core of radius $R_{\ast}$ is opaque, i.e.\ emission zones with
$z_i < 0$ and $p_i < R_{\ast}$ are obscured and produce no emergent
intensity.

Furthermore, we assume that the radial range between $R_\ast$ and
$r_{\rm sh}^{\rm min}$ is filled with homogeneous gas.  Thus
the ray from an emitter at $(p_i, z_i)$ towards the observer collects
an additional optical depth
\begin{equation}
\tau_{\rm h}^{\rm int} = \frac {\kappa \dot{M}}{4\pi} 
\int_{z_0}^{z^{\rm min}_{\rm sh}}
\frac{{\rm d}z} {r^2 v(r)}
\label{eq:taububble}
\end{equation}
when traveling though this internal sphere (cf.\
Eq.\,(\ref{eq:tauhz})). Here $z^{\rm min}_{\rm sh} = \sqrt{(r_{\rm
sh}^{\rm min})^2-p_i^2}$ and $z_0 = \max(z_i, -z_0)$.

Moreover, for radii larger than $r_{\rm sh}^{\rm max}$ the
fragments decayed and the wind is assumed to be
homogeneous again. The emission from all zones which lie further
inward suffers an additional absorption
\begin{equation}
\tau_{\rm h}^{\rm out} = \frac{\kappa \dot{M}}{4\pi} 
\int_{z_{\rm sh}^{\rm max}}^\infty
\frac{{\rm d}z}{r^2 v(r)}
\label{eq:tauinf0}
\end{equation}
with $z^{\rm max}_{\rm sh} = \sqrt{(r_{\rm sh}^{\rm max})^2 - p_i^2}$.
Since we generally assume that $r_{\rm sh}^{\rm max}$ is large enough
for the velocity at these distances being constant,
Eq.\,(\ref{eq:tauinf0}) becomes
\begin{equation}
\tau_{\rm h}^{\rm out} = \frac{\kappa \dot{M}}{4\pi p}
\left[\frac{\pi}{2}- \tan^{-1}\frac{z}{p}\right]\ .
\label{eq:tauinf}
\end{equation}

Therefore the optical depth along a ray between the emission zone $i$
and the observer is, in general, composed of the sum over the crossed
shell fragments, $\tau_i$, plus contributions from the homogeneous
parts of the wind inside and outside of the fragmented zone, i.e.\
$\tau_{\rm h}^{\rm int}$ and $\tau_{\rm h}^{\rm out}$,
respectively. Although this is fully implemented in our code, in the
discussion of this paper we will refer merely to $\tau_i$ to keep the
notations simple. With $\tau_i$ from now to be understood as the total
optical depth between the X-ray emitting zone $i$ and the observer,
the emergent intensity becomes
\begin{equation}
\label{eq:I-abs}
I^+_{\nu,i} (\nu,p,\varphi) = I_{\nu,i} (\nu,p,\varphi)\ \exp(-\tau_i)\ .    
\end{equation}

The emergent astrophysical flux, at some reference radius $R$, is
obtained by averaging the intensity over the disk of radius $R$,
\begin{equation}
F_\nu = \frac{1}{\pi R^2}\ \int_0^R {\rm d}p\ 
\int_0^{2\pi} p\ {\rm d}\varphi\ I^+_{\nu,i} (\nu,p,\varphi)\  .
\end{equation}
We insert $I^+_{\nu,i}$ from Eqs.\,(\ref{eq:I-em}) and
(\ref{eq:I-abs}), and perform the integral over the coordinates. Due
to the delta functions, this results in summing up the contributions
from all emitting zones,
\begin{equation}
\label{eq:F_nu}
F_\nu = \frac{1}{4\pi^2 R^2}\ \sum_{i=1}^{i_{\rm max}} L_i \exp(-\tau_i)\ 
\delta(\nu - \nu_{i,{\rm obs}})\    .
\end{equation}
We can calculate now the total X-ray line luminosity emerging from the
star. For this purpose, one has to multiply the physical flux $\pi
F_\nu$ with the surface of the reference sphere ($4\pi R^2$) and to
integrate over all frequencies, giving
\begin{equation}
L_{\rm X} = \sum_{i=1}^{i_{\rm max}} L_i\ \exp(-\tau_i)\     .
\end{equation}
Note that the internally released X-ray line luminosity $L_{\rm
X,int}$ from Eq.\,(\ref{eq:L_X}) is correctly reproduced by
Eq.\,(\ref{eq:F_nu}) in case of zero absorption.

Due to the discrete character of the emitting parcels of gas, the
emergent flux is composed of delta-function spikes at discrete
frequencies. A spectrum of finite resolution $\Delta \nu$ is obtained by
a corresponding binning. For each frequency index $k$, we compile a list
${\mathcal I}_k$ of those emission zones indices $i$ for which $\nu_{i,{\rm obs}}$
falls into the frequency interval $(\nu_k \!-\! \frac{1}{2}\Delta\nu,
\nu_k \!+\! \frac{1}{2}\Delta\nu)$. Then, by integrating the flux 
(Eq.\,(\ref{eq:F_nu})) over the frequency bin $k$, the binned flux at 
frequency $\nu_k$ becomes
\begin{equation}
\label{eq:F-bin}
F(\nu_k)\ = \frac{1}{4\pi^2 R^2 \Delta\nu}\ 
\sum_{i \in {\mathcal I}_k} L_i \exp(-\tau_i)\   .
\end{equation}
The emergent flux scales linearly with the internally released X-ray
line luminosity, $L_{\rm X,int}$ (cf.\ Eq.\,(\ref{eq:L_X})). It is
illustrative to normalize the emergent flux profile with respect to the
unabsorbed case. The total energy flux in the line for the unabsorbed
case, obtained by setting all $\tau_i = 0$ and integrating the previous
equation over frequency, is $L_{\rm X,int}\ / (4\pi^2 R^2)$.
Normalizing the emergent flux profile to this value, 
Eq.\,(\ref{eq:F-bin}) becomes
\begin{equation}
F^\ast(\nu_k)\ = \frac{1}{\Delta\nu\ L_{\rm X,int}}\ 
\sum_{i \in {\mathcal I}_k} L_i \exp(-\tau_i)\      .
\label{eq:lineprofl}
\end{equation}
For convenience we will display the computed line profiles as function
of dimensionless frequency $x$, measured relative to the line center
and in Doppler units referring to the terminal wind velocity
$v_\infty$,
\begin{equation}
x = \frac{\nu - \nu_0}{\nu_0\ v_\infty / c}\ . 
\end{equation}
Hence the line profile is confined between normalized frequencies $-1$
and 1.  Using frequency bins $\Delta x$, the normalized flux finally
becomes
\begin{equation}
F^\ast(x_k)\ = \frac{1}{\Delta x\ L_{\rm X,int}}\ 
\sum_{i \in {\mathcal I}_k} L_i \exp(-\tau_i)\      .
\label{eq:fluxf}
\end{equation}
If the emitting zones propagate with constant velocity and
there is no absorption, the resulting line profile is box-shaped with
$F^\ast(x_k)=0.5$ for $|x_k| < 1$, so that the integral over the whole
profile correctly contains all internally released X-ray line
luminosity (cf.\ Eq.\,(\ref{eq:L_X})).

Equation\,(\ref{eq:fluxf}) yields the line profile for one specific
random trial of the emitting spots and absorbing fragments. A
different random trial will result in a different line profile, and a
large number of trials (a few thousands, typically) is required to
achieve a good statistical average. Phenomenologically, we can
consider each trial as a snapshot of the wind. Typical exposure times
of an X-ray observation are much larger than the dynamical time scale
$R_\ast/v_\infty$ of the stellar wind, which corresponds to an average
over many snapshots. Moreover, our model is 2-D, whereas in a 3-D wind
different azimuthal angles will display statistically independent
patterns of emission and absorption, whose different line profiles are
also averaged in a flux measurement. Investigating the variability of
X-ray lines would be interesting, but is beyond the scope of the
present paper.

\section{Illustration of basic effects}
\label{sect:basic-effects}

In this section we concentrate on simplified cases of wind emission
and absorption to demonstrate the basic effects of fragmentation on
line profiles.

\begin{figure}[hbtp] 
\epsfxsize=\columnwidth
\centering \mbox{\epsffile{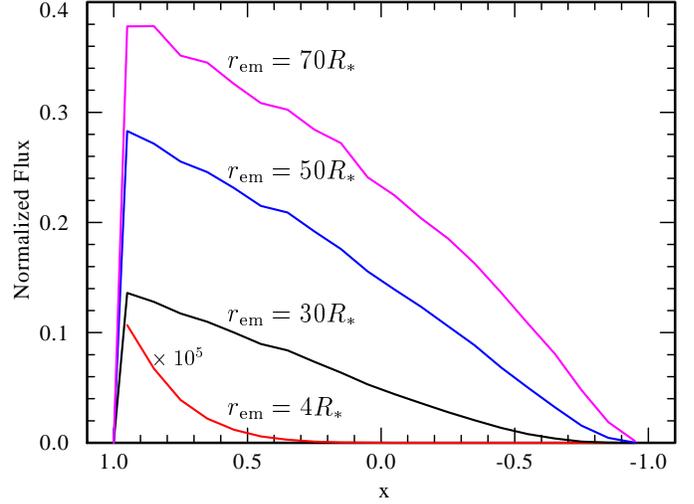}} 

\caption {Emission line profiles from a homogeneous wind of
constant velocity. The radial optical depth of the photosphere is
$\tau_\ast = 50$ for all shown profiles. All hot gas emitting the X-ray
line is assumed to be located at the same radius $r_{\rm em}$, which
is varied for the different profiles shown (see labels). All profiles
appear skewed and blueshifted. The lowest curve ($ r_{\rm em} =
4\,R_\ast$) is zoomed $10^5$ times.}      

\label{fig:sm} 
\end{figure}

\medskip\noindent {\bf Homogeneous wind.} The line profiles from a
homogeneous outflow are widely discussed in the literature (e.g.\
MacFarlane et al.\ \cite{MacFarlane91}).  Figure\,\ref{fig:sm}
displays the line profiles for an optically thick, homogeneous wind of
constant velocity. All emission zones are placed at the same radius,
$r_i = r_{\rm em}$, which is varied for the different profiles shown
in the figure (labels). The further out the emission is located, the
less absorption occurs along rays in outward direction, i.e.\ in the
blue part of the line profile. On the other hand, radiation from the
back hemisphere which forms the red part of the line profile must
always penetrate the whole wind. Therefore the resulting profiles
appear skewed and blueshifted. Note that the lowest curve,
corresponding to an emission radius of $4\,R_\ast$, is zoomed $10^5$
times, which means that essentially all radiation formed deep in the
wind is totally absorbed.

\begin{figure}[hbtp] 
\epsfxsize=\columnwidth
\centering \mbox{\epsffile{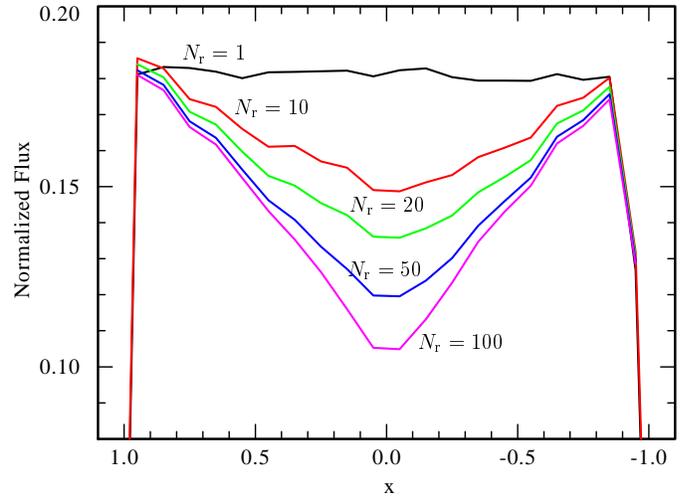}} 

\caption{Line profiles from a wind where the absorbing material is
compressed in $N_r$ concentric shells distributed between $r_{\rm
sh}^{\rm min} = 4\,R_\ast$ and $r_{\rm sh}^{\rm max} =
100\,R_\ast$. The emission originates at $r_{\rm em} = 4\,R\ast$, the
wind velocity is constant ($v=\vinf$) and the total radial optical
depth across all shells is $\tau_\ast = 1$. The homogeneous wind below
$4\,R_\ast$ is omitted here. The different profiles shown correspond
to different numbers $N_r$ of concentric spheres (labels), whose
randomly chosen radii are all larger than that of the emission zone.}

\label{fig:circtau1cv}
\end{figure}

\medskip\noindent {\bf Concentric spheres.}
Figure\,\ref{fig:circtau1cv} shows line profiles from the concentric
spheres model, omitting the homogeneous inner and outer parts of the
wind. Radiation emitted close to the stellar core is attenuated by an
ensemble of concentric spheres located above the emission site. The
location of the spheres is determined by Eq.\,(\ref{eq:dsh}). The
optical depth of a shell crossed by the ray between an emitter and the
observer,$\tau_j$, scales inversely with $|\mu|$
(Eq.\,(\ref{eq:tsh})), i.e.\ depends on the intersection angle between
ray and shell, and becomes large when the shell is met under a small
angle. Therefore the shape of the line profile is determined by the
range of angles under which the absorbing shell can be crossed by an
emitted photon. For emitting spots placed at $r_i$ and absorbing
fragments at $r_j$, the smallest possible value of $|\mu|$ is $\sqrt{1
- (r_i/r_j)^2}$. Consider, for instance, the case of just one
concentric shell ($N_r=1$) which has swept up all material between
$r_{\rm sh}^{\rm min}$ and $r_{\rm sh}^{\rm max}$, moving with
constant velocity. According to Eq.\,(\ref{eq:dsh}), the shell is
located at $\sqrt{r_{\rm sh}^{\rm min} \cdot r_{\rm sh}^{\rm max}}$ in
order to conserve the radial optical depth. For the example shown in
Fig.\,\ref{fig:circtau1cv}, $r_{\rm sh}^{\rm min} = 4\,R_\ast$ and
$r_{\rm sh}^{\rm max} = 100\,R_\ast$, and hence the shell is located
at $r_j = 20\,R_{\ast}$. The emission is located at $r_i =
4\,R_{\ast}$, i.e.\ at a much smaller radius. Therefore the minimum
value of $|\mu|$ is close to unity, and the optical depth is nearly
the same for all rays crossing the shell.  Consequently, in this
example with only one shell, the line profile is nearly flat-topped, the
normalized flux being $F^*(x) = 0.5\,\exp(-\tau_j) \approx
0.5/e$.

We increase now the number of shells $N_r$ into which the material is
compressed. According to their random distribution, some of the shells
lie closer to the minimum fragmentation radius than the one shell in
the $N_r = 1$ example. The larger $N_r$, the nearer is the innermost
shell to the emitting sphere. For rays tangential to the emitting
sphere, i.e.\ at line center, the intersection angle with the
innermost shell becomes small. These rays experience strong
absorption, which leads to a depression in the line center.

\begin{figure}[hbtp] 
\epsfxsize=\columnwidth
\centering 
\mbox{\epsffile{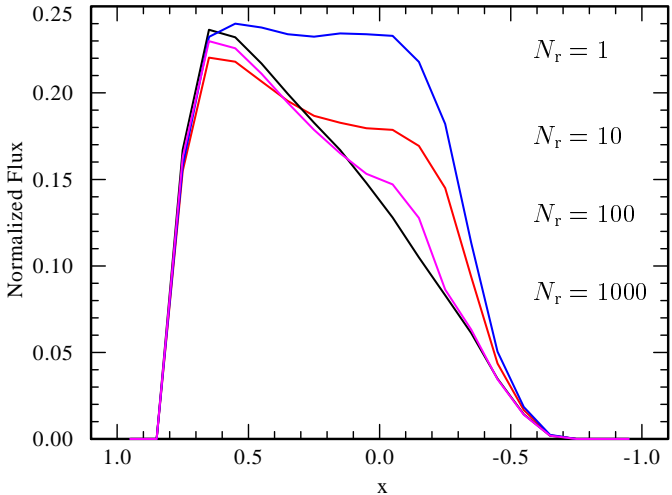}} 

\caption{The effect of a homogeneous inner wind. All parameters are
the same as in Fig.\,\ref{fig:circtau1cv}, except that the central
part of the wind between $R_\ast$ and $r_{\rm em}$ is now filled with
homogeneous absorbing material. When the number of concentric spheres,
$N_r$, becomes large, the emergent line profile converges to the line
profile from a homogeneous wind of the same $\tau_\ast$. The latter is
shown by the thick curve, which is indistinguishable from the case of
$N_r=1000$ shells.}

\label{fig:circtau1}
\end{figure}

However, these profiles are not realistic because we have artificially
omitted any absorption inside a radius of $4\,R_\ast$. We now fill
this region with the homogeneous inner part of the wind, while all
other parameters remain as in Fig.\,\ref{fig:circtau1cv}. The effect
is demonstrated in Fig.\,\ref{fig:circtau1}. If there is absorbing
material between the stellar core and the X-ray emitting spherical
layer, then radiation from the receding back hemisphere has to pass
through it, whereas radiation from approaching part has
not. Therefore, the red part of the line, $x>0$, is always depleted
compared to the blue part, $x<0$. The blue part is flat-topped for a
small number of shells, but with increasing $N_r$ the resulting
profiles approach the shape from a homogeneous wind. This result does
not depend on whether the central part of the wind is homogeneous, or
arranged in a number of concentric shells.

\begin{figure*}[hbtp] 
\epsfxsize=.9\textwidth 
\centering 
\mbox{\epsffile{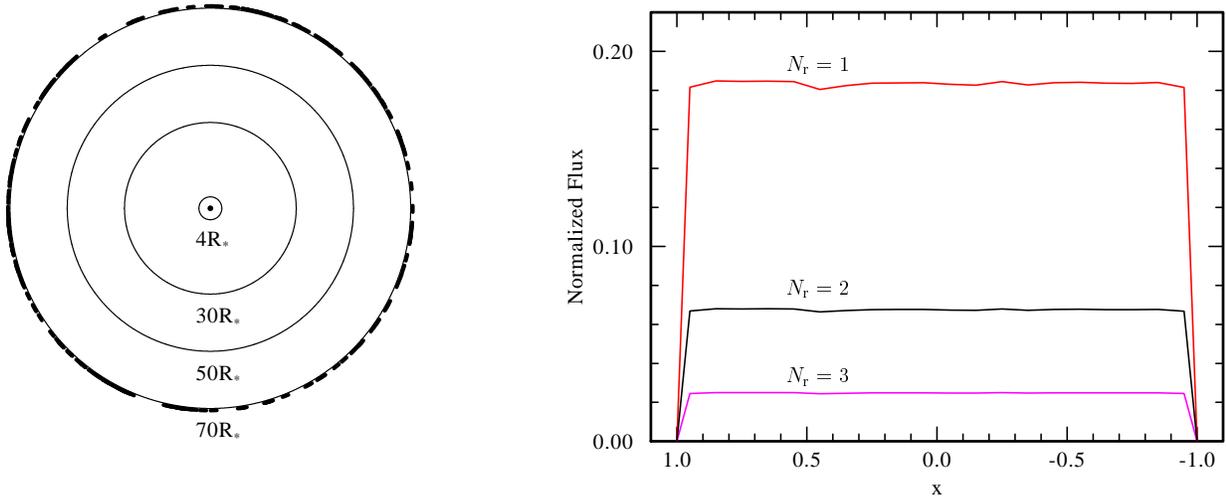}}

\caption{The broken sphere model. One opaque sphere ($\tau_\ast =
4$) is divided into $N_\vartheta$ fragments, which are then randomly
redistributed in angle. One of the random trials is shown in the left
panel. Constant velocity is assumed. Emission arises from one sphere,
for which different radii $r_{\rm em}/R_\ast =$ 4, 30, 50 and 70 are
tested, but the emergent line profiles are independent from this
choice. The line profiles shown in the right panel are obtained after
averaging over $\approx 10^3$ random realizations, for the case that
$N_r =$ 1, 2 or 3 spheres were broken into fragments (labels).}

\label{fig:oneshell_b}
\end{figure*}

\medskip
\noindent{\bf Optically thick broken spheres.} Let us start with just
one absorbing shell of large optical depth, assuming that there
is no absorption in the inner part of the wind, below the emission
site.  When all emission is formed inside absorbing shell, it will
be completely trapped and the external observer will not register any
emergent radiation.

In the broken sphere model, we break this absorbing shell into
$N_{\vartheta}$ fragments, for simplicity all having the same size. In
the 2-D cut each fragment is a fraction of a circle, i.e.\ an arc.
These fragments are placed randomly on the circle, which means that
the coordinate $\vartheta_j$ of the center of each fragment is a
random number chosen uniformly between 0 and 2$\pi$.

Clearly, the randomly distributed fragments will overlap in some
directions, and leave gaps in others. Remember that they covered one
full sphere before the latter was broken. The radiation from inside
the sphere can pass through these gaps entirely unattenuated.

The left panel of Fig.\,\ref{fig:oneshell_b} shows one random
distribution of fragments. All emission is coming from just one sphere
with radius $r_{\rm em}$, for which different values are tested
(labels). Constant velocity is assumed. Therefore the cosine of the
angle between the direction to the observer and to an emitting spot is
the normalized frequency, $\mu_i= x$, at which the observer receives
the line photons from this spot. For one random distribution of the
absorbing fragments, the optical depth along the line of sight from
one specific emission spot at ($\mu_i, r_i$) towards the observer is
either zero, or big.

What is the probability that $\tau_i (\mu_i, r_i)=0$? For solid
spheres, the probability that a gap occurs along a line of sight is
nil. For broken spheres, simple combinatoric considerations show that,
on average, a fraction $\exp(-1)$ of the sphere is free from absorbing
fragments.

When a photon has a probability of $\exp(-1)$ to find a gap in the
broken sphere, the emergent intensity is $I^+ = I_0\exp(-1)$, always
understood in the statistical limit of averaging over many random
settings. In other words, the optical depth along the line of sight is
unity. Note that this holds for all line frequencies, because the
fractional area of gaps in a surface does not depend on the projection
angle under which the surface is seen. Thus the emergent line profile
is flat-topped, independent from the radius $r_{\rm em}$ of the
emitting sphere (cf.\ Fig.\,\ref{fig:oneshell_b}, right panel).

It is important to realize that the flat-topped shape is due
to two conditions. First, the emission must be produced in a narrow
range of velocities, in order that the un-absorbed profile would be
flat-topped.  Second, the absorbing fragments must be thin, i.e.\
their radial extent must be small compared to their lateral
size. Globular clouds, for instance, would not meet this condition.  

So far, we considered one single broken sphere. What if there are
$N_r$ concentric broken spheres, all located above the emission
forming region? Since the $\vartheta$-randomization of fragments is
independent for each sphere, the optical depth for each of them is
unity, and the total optical depth adds up to $\tau_i=N_r$.

The right panel of Fig.\,\ref{fig:oneshell_b} shows the line profiles
obtained for $N_r = 1$, 2 and 3. The left panel shows one realization
of a random distribution of absorbing fragments, whereas the line
profiles are accumulated over a large number of trials.

The emergent flux profiles are flat-topped. Compared to the case
without any absorption ($F^{\ast}(x) = 0.5$), the flux is reduced by a
factor $\exp(-N_r)$, as expected from our statistical
considerations. The key difference between the line profiles in
Figs.\,\ref{fig:circtau1} and \,\ref{fig:oneshell_b} is due to the
fact that in the former case the transport of radiation is {\em
through the material}, which leads to an angular-dependent optical
depth, while in the latter case the radiation emerges {\em through the
empty gaps}, where the probability to find a gap does not depend on
angle.

\medskip \noindent {\bf Optically thick random fragments}. So far we
treated first the case where the radius of the absorbing (full)
concentric shells was chosen randomly; and second the case of one or
more shells which were broken into fragments and distributed randomly
in angle, while the fragments kept their given radial coordinate.  We
combine these cases now and distribute the individual fragments
randomly and independently in both the angular and the radial
coordinate. For the moment the fragments are assumed to be opaque.

\begin{figure*}[hbtp] 
\centering 
\epsfxsize=.9\textwidth 
\mbox{\epsffile{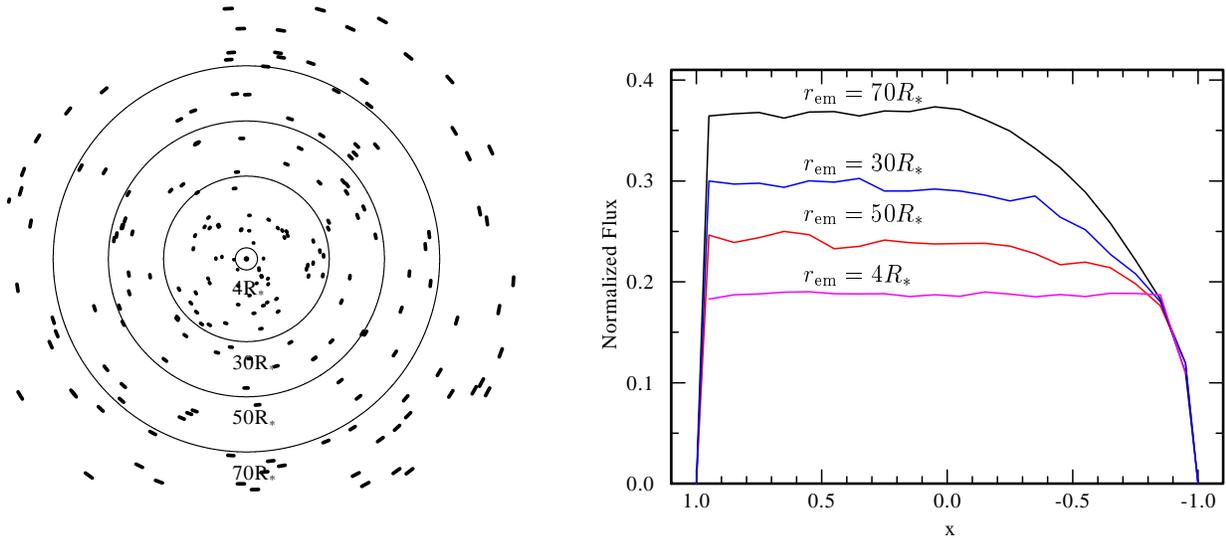}} 

\caption{Random fragments model. Left panel: one random trial of the
model where the absorbing shell fragments are randomly distributed in
radius (uniformly between 4.01 and 100\,$R_\ast$) and in angle. On
average there is one fragment along each radial direction, $\langle
N_r \rangle = 1$. The fragments in this example are opaque, and the
central part of the wind ($r < 4\,R_\ast$) is void. Right panel: line
profiles, for emission located at $r_{\rm em}/R_\ast$ = 4, 30, 50 and
70 (labels). For $r_{\rm em} = 4\,R_\ast$, all absorbers are outside
of the emission site and the resulting profile is flat-topped as in
the broken sphere model (cf.\ Fig.\,\ref{fig:oneshell_b}). If emission
is located within the absorbing wind, as in cases with $r_{\rm
em}/R_\ast$ = 30, 50 and 70, the red part of the profile is
additionally absorbed in the back hemisphere, while the blue part of
the line profile remains flat-topped.}

\label{fig:oneshell_c}
\end{figure*}

The left panel of Fig.\,\ref{fig:oneshell_c} shows one trial for this
random fragments model. The total solid angle of all fragments
corresponds to one full sphere. The absorbing fragments are
distributed between 4.01 and 100\,$R_\ast$.

To start with, let us assume that the emitters are located closer to
the star than {\it any} absorbing material. In
Fig.\,\ref{fig:oneshell_c} this corresponds to the example with
$r_{\rm em} = 4\,R_\ast$. According to the parameters of this model,
there is, after averaging over many trials, one absorbing fragment in
each radial direction, $\langle N_r \rangle = 1$.  This is exactly the
same situation as in the broken sphere model considered
above. Emerging radiation suffers absorption with an optical depth
$\tau_i=1$, leading to the formation of a rectangular line profile.

Next let us consider emission which is permeated with the absorbing
fragments. Again for clarity, we assume that the radiation is produced
at a single radius $r_{\rm em}$. In Fig.\,\ref{fig:oneshell_c} the
radii of emission are chosen to be 4, 30, 50, and 70\,$R_\ast$. When
the emission takes place in outer parts of the wind, say at
$70\,R_\ast$, there are not many fragments left between $r_{\rm em} =
70\,R_\ast$ and $r_{\rm sh}^{\rm max} = 100\,R_\ast$. As can be seen
from the figure, there are many gaps for photons to escape
unattenuated from the front hemisphere towards the observer. Since we
assume constant velocity in this example, the fragments are uniformly
distributed in radial direction. Hence the average number of absorbing
fragments along a radial ray from $r_{\rm em}$ to $r_{\rm sh}^{\rm
max}$ is $\langle N_r \rangle (r_{\rm sh}^{\rm max} - r_{\rm
em})/(r_{\rm sh}^{\rm max} - r_{\rm sh}^{\rm min})$, with $\langle N_r
\rangle = 1$ in our example. For instance, with $70\,R_\ast$ for the
radius of emission and $100\,R_\ast$ for the outer boundary, the
average number of absorbing fragments in radial direction is about
0.3.

This number plays the role of an {\em effective optical depth} $\bar
\tau$; the probability that there is {\em no} fragment along the
radial ray from $r_{\rm em}$ to the observer is $\exp(-\bar
\tau)$. This concept will be discussed on a more general basis in
Sect.\,\ref{sect:effective_opacity}.  See also the discussion after
Eq.\ (3) in Paper I.

\begin{figure*}[hbtp] 
\centering 
\epsfxsize=.9\textwidth 
\mbox{\epsffile{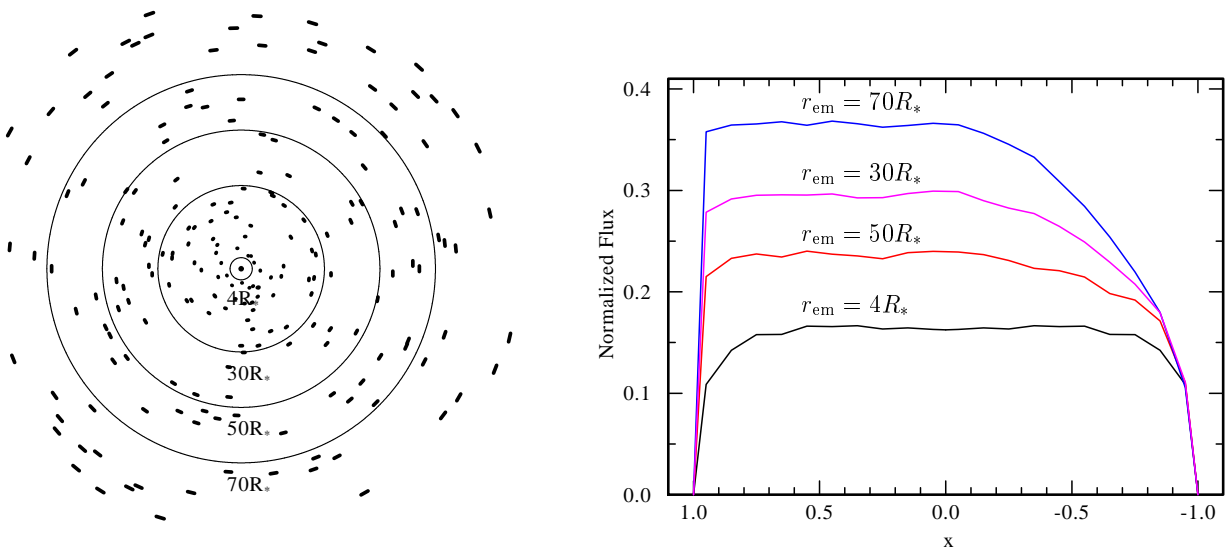}} 

\caption{Cones model. An absorbing sphere (i.e.\ a circle in the 2-D
cross section) is divided into 200 equal arcs, which are randomly
distributed in radius. In contrast to the random fragments model in
Fig.\,\ref{fig:oneshell_c}, the fragments are kept in their angular
position (``cone'').  Otherwise, the model is the same as in
Fig.\,\ref{fig:oneshell_c}. By eye one can not catch a big difference
between this distribution and the one for the random fragments
model. This is reflected by the line profiles shown in the right
panel, which look very similar to those in
Fig.\,\ref{fig:oneshell_c}. The overall flux is only slightly lower
than in case of random fragments. The biggest difference is at the
blue edge of the profile. This is due to the fact that in the cones
model the radial optical depth is preserved (here, $\tau_\ast = 4$),
and photons emitted in almost radial direction towards the observer
cannot find gaps to escape.}
\label{fig:oneshell_d}
 \end{figure*}

For the interpretation of Fig.\,\ref{fig:oneshell_c} we must
understand how this effective optical depth depends on the value $\mu$
of the emitting spot located on the sphere of radius $r_{\rm em}$
(remember that the radiation from $\mu$ is observed at dimensionless
frequency $x = \mu$). Obviously, the probability for a ray to hit a
fragment depends only on the radial interval which the ray has to
cross. For the front hemisphere ($\mu > 0$) this range is $r_{\rm
sh}^{\rm max}-r_{\rm em}$, independent of $\mu$. Therefore the blue
part of the profiles in Fig.\,\ref{fig:oneshell_c} is flat-topped, and
the flux level is $\exp(-\bar\tau)$ times the the flux of the
un-absorbed emission.

If the emitters are located at $r_{\rm em} = 4\,R_\ast$, the
flat-topped shape extends also to the red side of the profile, because
of our somewhat artificial assumption in this example that the central
part of the wind ($r < 4\,R_\ast$) is void. However, when the emitting
spots are permeated with the fragments (say, at $r_{\rm em} =
50\,R_\ast$), a ray from an emitting spot on the back hemisphere
($\mu<0)$ with coordinates $(p_i, z_i < 0)$ collects optical depth
from both hemispheres. In radius, the ray first descends from $r_{\rm
em}$ to $r=p_i$ (back hemisphere), and than climbs from $r=p_i$ to
$r_{\rm sh}^{\rm max}$ (front hemisphere). Thus the total radial range
crossed is $r_{\rm sh}^{\rm max}+r_{\rm em}-2p_i$, and the optical
depth is proportional to this range. Therefore the red part of the
line becomes strongly absorbed and skewed.

\medskip \noindent{\bf Cones.} This model was introduced in
Sect.\,\ref{sect:distribution_of_absorbers}. Each absorbing sphere is
divided into $N_\vartheta$ segments (i.e.\ arcs in the two-dimensional
cross section), which are randomly distributed in radius. In contrast
to the random fragments model, the fragments are kept at their angular
position (``cone''), and not randomized in lateral direction. This
model is interesting, because the radial optical depth and the mass
are conserved for each radial direction. One of the random trials for
the distribution of absorbing fragments is shown in the left panel of
Fig.\,\ref{fig:oneshell_d}. As in the previous examples, the emission
is located at a unique radius $r_{\rm em}$, and there is no
absorption below this radius. As seen by eye, this distribution of
the fragments can hardly be distinguished from the one in
Fig.\,\ref{fig:oneshell_c}.

The resulting line profiles are shown in the right panel of
Fig.\,\ref{fig:oneshell_d}. They also do not differ much from the line
profiles obtained in the random fragment case. The overall flux is
only slightly lower. The biggest difference is at the blue edge of the
profiles. This is due to the fact that in the cones model the radial
optical depth is conserved (here, $\tau_\ast = 4$), and photons
emitted in almost radial direction towards the observer cannot find
gaps to escape from absorption. But apart from the blue edge, the
radial randomization of the absorbing fragments provides escape
channels for the photons, acting in the same way as in the random
fragments model.

It order to verify this numerical results for 3-D, we made
numerical experiments to illustrate the effect of random radial
re-positioning of fragments of a sphere. We cover the unit sphere with
spherical triangles of similar size, mimicking absorber
fragments. Spherical coordinate lines are not appropriate here, due to
the dependence of solid angle on $\sin\vartheta$. The sphere is
divided into octants. For each of the three grand circle segments
delimiting an octant surface, the midpoint at half arc length is
connected to the midpoints of the other two grand circle
segments. This divides the octant surface into four spherical
triangles.  For each of them this procedure (division of grand circle
segments at half arc length, and connection to nearest neighbors) is
repeated.  After six such subdivisions one obtains $4^7 =16384$
triangles on the front hemisphere. These triangles are of similar size
and shape, but not exactly uniform.

\begin{figure}[bthp]
\centering 
\epsfig{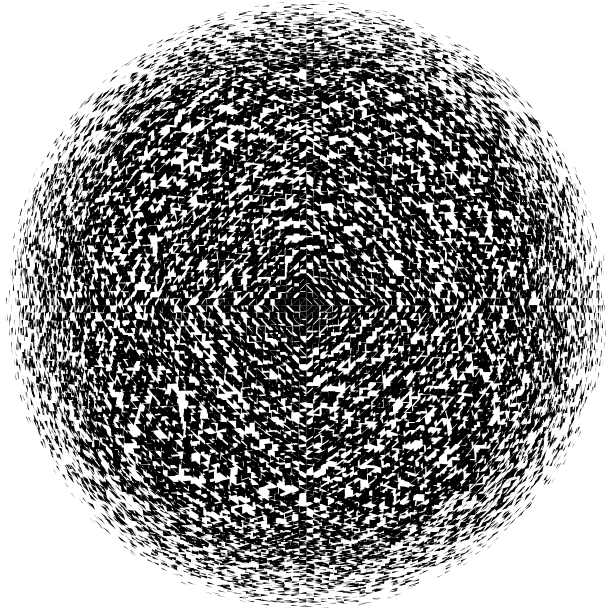}
\caption{Sky projection of radially randomized fragments. The front
hemisphere of a unit sphere is cut into 16\,384 roughly equal
spherical triangles, which are subsequently randomly redistributed
between $r$ = 0.8 and 1.0 while their angular position is kept.}
\label{fig_af2}
\end{figure}

\begin{figure}[hbtp]
\centering 
\epsfxsize=7.5cm
\mbox{\epsffile{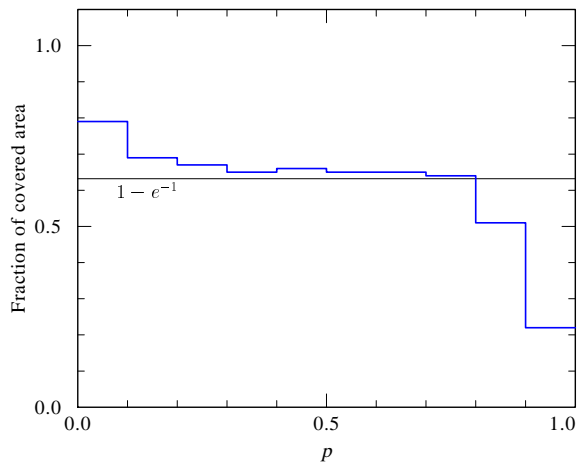}} 

\caption{Fraction of projected surface covered by radially randomized
fragments.  The sky projection of the fragmented hemisphere from
Fig.\,\ref{fig_af2} is divided into 10 concentric rings of equal width
in impact parameter $p$, and the fraction of covered (black) surface
is determined numerically. Except for the center and the limb region,
the coverage is $\approx 1-e^{-1}$, i.e.\ close to the value expected
if the fragments were randomly distributed in their {\em angular}
position, but kept on the unit sphere.}
 
\label{fig:fig_af3}
\end{figure}

Next, the radial location of these (opaque) triangles is chosen
randomly and uniformly between $r=0.8$ and 1. The corresponding sky
projection is shown in Fig.~\ref{fig_af2}. In fact, the distribution
appears random and uniform, except in the disc center and close to the
limb. To quantify this, we divide the projected disk into 10 rings of
equal radial extent and determine numerically the fraction of the ring
area which is covered by at least one fragment
(Fig.\,\ref{fig:fig_af3}). The complement is the fractional area which
remains free of any projected fragment, while fragments are
overlapping at other locations. For central rays $p\ll 1$, radial
randomization cannot move fragments out of the line of sight. Hence,
the central region $r\le 0.2$ remains largely obscured (small fragment
overlap), $\tau>1$. For $0.2\le r\le 0.8$, the transmission is close
to $1 - e^{-1}$, as expected if the fragments were randomly
distributed {\em laterally} on the surface of the sphere. For $r>0.8$
finally, $\tau < 1$ because fragments are radially distributed over
this range. To summarize, for an observer at infinity, radial
randomization in a spherically symmetric wind is largely
indistinguishable from lateral randomization.

\section{Effective opacity}
\label{sect:effective_opacity}

An alternative approach to the stochastic numerical modeling of this
paper is to introduce an {\em effective} optical depth $\bar\tau_i$
for a random distribution of absorbing fragments by statistical means
(cf.\ Paper\,I). Both the numerical and the analytical approach should
yield identical results for the same parameters. In this
section we concentrate on the analytical description, and compare the
resulting line profiles with those obtained in the numerical
simulations.

Let us define the effective optical depth $\bar\tau_i$ in the
conventional way, as integral over an effective opacity $\alpha$
(dimension: per length) along the path,
\begin{equation}
\bar\tau_i\,=\,\int_{z_i}^{z_{\rm sh}^{\rm max}} \alpha\ {\rm d}z\ .
\label{eq:defal}
\end{equation}
The radiative transfer equation keeps its usual form
\be
dI_{\nu}\,=\,-\alpha\,I_\nu {\rm d}z,
\label{eq:al}
\ee
with the well-known solution Eq.\,(\ref{eq:I-abs}).

The effective opacity is the product of the number density of
absorbers and the (average) area of the absorbing fragments projected
on a plane perpendicular to the line of sight.  Moreover, we allow (in
contrast to the usual definition of atomic cross sections) for
absorbers that are not completely opaque, but have a probability,
${\cal P}$, that a photon hitting the cross section is absorbed.

The average number of shells within a unit radius is termed $n(r)$. In
case of constant velocity, $n(r) = \langle N_r \rangle /(r_{\rm
sh}^{\rm max}-r_{\rm sh}^{\rm min})$ is also constant. With variable
velocity, the radial density of shells scales with $v^{-1}$ (cf.\
Sect.\,\ref{sect:distribution_of_absorbers}). Therefore we have
\be
n(r) = \frac{n_0}{v(r)}\ ,
\label{eq:def_n(r)}
\ee
with
\be
n_0 = \langle N_r \rangle \left/ 
\int^{r_{\rm sh}^{\rm max}}_{r_{\rm sh}^{\rm min}} \frac{{\rm d}r}{v(r)}\ 
\right .
\ee

\begin{figure}[thbp] 
\centering 
\epsfxsize=8cm
\mbox{\epsffile{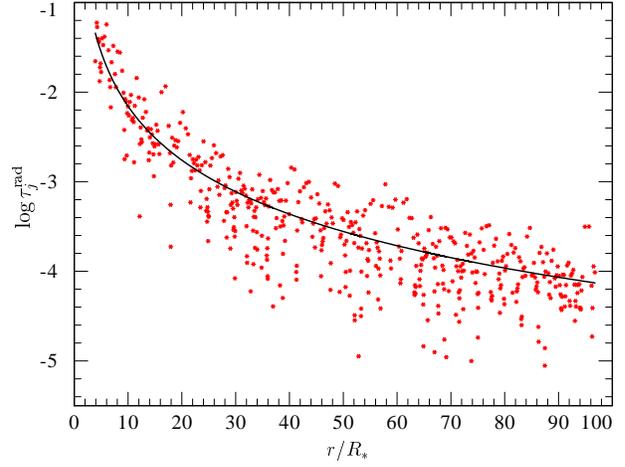}} 

\caption{Radial optical depth of individual fragments. For one
selected cone in the cones model, the total radial optical depth
$\tau_\ast = 1$ splits into $N_r = 500$ shells located at random radii
$r_j$. The optical depth $\tau_j^{\rm rad}$ of each fragment varies
according to the random algorithm used to define the mass in each
shell, and is denoted by dots. The continuous line gives the average
radial optical depth $\bar \tau_j^{\rm rad}$ of the shells, if the
mass is uniformly distributed over all fragments. Further parameters
of the model are $r_{\rm sh}^{\rm min} = 4\,R_\ast$, $r_{\rm sh}^{\rm
max} = 100\,R_\ast$, and $\beta = 1$ for the velocity-law exponent.}

\label{fig:tau-r}
\end{figure}

As can be seen from Fig.\,\ref{fig:coord}, the area of a fragment
projected to the plane perpendicular to the line of sight scales with
$|\mu|$. The total area of all absorbers within a spherical layer of
radius $r$ and thickness ${\rm d}r$ is $n(r)\ |\mu|\ 4\pi r^2 {\rm
d}r$, while the volume itself is $4\pi r^2 {\rm d}r$.  Thus the
effective opacity is
\be 
\alpha(r,\mu) = n(r)\ |\mu|\ {\cal P}\ .
\label{eq:alpha}
\ee
Note that this opacity is not isotropic as usual, but angular-dependent. 
The probability ${\cal P}$ that a photon which encounters a fragment is  
absorbed is 
\be 
{\cal P}=1-{\rm e}^{-\tau_j}\ ,
\label{eq:calp}
\ee
i.e.\ is a function of the optical depth $\tau_j$ of fragments at $r$
with orientation $\mu$ as given by Eq.\,(\ref{eq:tsh}).

As long as one avoids to integrate through the point where $\mu = 0$,
the path element along the ray, ${\rm d}z$, can be substituted by
${\rm d}r$, using
\be
{\rm d}z = \mu^{-1}\,{\rm d}r\ .
\label{eq:path}
\ee 
By inserting Eqs.\,(\ref{eq:alpha}), (\ref{eq:calp}) and (\ref{eq:path})
into Eq.\,(\ref{eq:defal}) we obtain for the effective optical depth 
between an emitting point at $(r_i, \mu_i)$ and the observer
\be
\bar\tau_i(r_i,\mu_i) = \int_{r_i}^{r_{\rm sh}^{\rm max}} 
n(r)\ (1-{\rm e}^{-\tau_j})\ {\rm d}r\ .
\label{eq:tint} 
\ee

Remarkably, in contrast to Eq.\ (\ref{eq:tauhz}) neither $\dot M$ nor
$\kappa$ appears explicitly in this expression for $\bar\tau_i$. The
effective optical depth is determined essentially by the geometrical
distribution of fragments, as explained earlier.

In the following we will consider the limits of optically thick and 
optically thin fragments, $\tau_j\gg 1$ and $\tau_j \ll 1$.

\subsection{Optically thick case ($\tau_j\gg 1$)}

If all cool material is assembled in a number of optically thick
fragments, the emergent radiation is only due to the presence of gaps
between the fragment. The radiation passes through these gaps entirely
unattenuated, but all photons which intersect a fragment are
absorbed. Therefore the probability ${\cal P}$ is unity. In case of
constant velocity, the integral in Eq.\,(\ref{eq:tint}) is taken over
the constant $n(r)$ and reduces to (see Eq.\, (4) in paper I)
\be
\bar\tau_i(r_i,\mu_i) = \frac{\langle N_r \rangle}
{r_{\rm sh}^{\rm max}-r_{\rm sh}^{\rm min}}
\left\{
\begin{array}{lcc} 
r_{\rm sh}^{\rm max}-r_i      &   &{\rm if}\,\,\mu_i > 0 \\
r_{\rm sh}^{\rm max}+r_i-2p_i &   &{\rm if}\,\,\mu_i < 0.
\end{array}
\right.
\label{eq:tauang}
\ee
The line profiles from our stochastic numerical model shown in
Fig.\,\ref{fig:oneshell_c} agree extremely well with this analytical
result.

\subsection{Optically thin case ($\tau_j\ll 1$)}

The optical depth of individual absorbers may be small because the
total optical depth of the wind, $\tau_\ast$, is small, or because the
optical depth is distributed over very many fragments (i.e.\ $N_r$ is
large). If the absorbing fragments are optically thin, the emergent
flux is both due to radiation which passed unattenuated through gaps,
{\it and} to radiation which crossed fragments and was only partly
absorbed.

For small optical depths, $\tau_j \ll 1$, the exponent in
Eq.\,(\ref{eq:calp}) can be expanded as $\exp(-\tau_j) \approx 1 - \tau_j$.
Consequently ${\cal P}\approx\tau_j$ and Eq.\,(\ref{eq:alpha})
simplifies to
\begin{equation}
\alpha(r,\mu) = n(r)\ |\mu|\ \tau_j\ .
\label{eq:smta}
\end{equation}
Expressing the directional optical depth of the fragment, $\tau_j$, by 
its radial optical depth $\tau_j^{\rm rad}$ (Eq.\,(\ref{eq:tsh})) yields 
\be
\alpha(r,\mu) = n(r)\ \tau_j^{\rm rad}\ .
\label{eq:thin_effective_opacity}
\ee
Therefore, in the limit of optically thin fragments the effective
opacity becomes isotropic, as the usual opacity in homogeneous winds.

We insert now for $\tau_j^{\rm rad}$ the {\em average} optical depth
of a fragment $\bar\tau_j(r)$ at radius $r$. Remember that in our
stochastic wind model the amount of matter swept up into an individual
fragment is a random number. This is illustrated in
Fig.\,\ref{fig:tau-r}, which shows for one cone of the cones model how
the total radial optical depth $\tau_\ast = 1$ splits into $N_r = 500$
shells located at their individual radii $r_j$. To define an average
optical depth, each shell shall have the same amount of
mass. According to the geometrical dilution, the radial optical depth
of the shells scales with $r^{-2}$, so that we can write
\be
\bar\tau_j(r) = \frac{\tau_0}{r^2}\ . 
\label{eq:tau-bar}
\ee
The radial integral over the effective opacity must yield the total 
optical depth $\tau_\ast$, i.e.\ 
\be
\int^{r_{\rm sh}^{\rm max}}_{r_{\rm sh}^{\rm min}}  n(r)\
\bar\tau_j(r)\ {\rm d}r = \tau_\ast\ , 
\ee 
which leads with Eq.\,(\ref{eq:def_n(r)}) to 
\be 
n_0\ \tau_0 \int^{r_{\rm sh}^{\rm max}}_{r_{\rm sh}^{\rm min}}  
\frac{{\rm d}r}{r^2 v(r)} = \tau_\ast = 
\frac{\kappa \dot M}{4\pi n_0} 
\int^{r_{\rm sh}^{\rm max}}_{r_{\rm sh}^{\rm min}}  
\frac{{\rm d}r}{r^2 v(r)}   
\ee 
where on the right-hand side we expressed the radial optical depth
using its original definition for a homogeneous wind. Thus we have
$\tau_0 = \kappa \dot{M}/(4 \pi n_0)$. The continuous line in
Fig.\,\ref{fig:tau-r} shows the run of the average optical thickness
of fragments as function of radius (Eq.\,(\ref{eq:tau-bar})). The
effective opacity in Eq.\,(\ref{eq:thin_effective_opacity}) for the
optically thin limit finally becomes
\be
\alpha(r,\mu) = \frac{n_0}{v(r)} \frac{\tau_0}{r^2} = 
\frac{\kappa \dot{M}}{4\pi} \frac{1}{r^2 v(r)} \ ,
\ee
which is identical to the opacity $\chi$ of a homogeneous wind given
by Eq.\,(\ref{eq:kappa}).

To summarize, we have demonstrated that if the optical depth across
all fragments is small, $\tau_j\ \ll 1 $, the effective opacity in the
fragmented wind is identical to the opacity of a homogeneous wind of
otherwise the same parameters. We may conclude that the
homogeneous wind is the limiting case of a fragmented wind, when the
radial number of fragments is big and/or the total optical depth is
small.

\begin{figure}[hbtp] 
\epsfxsize=\columnwidth
\centering 
\mbox{\epsffile{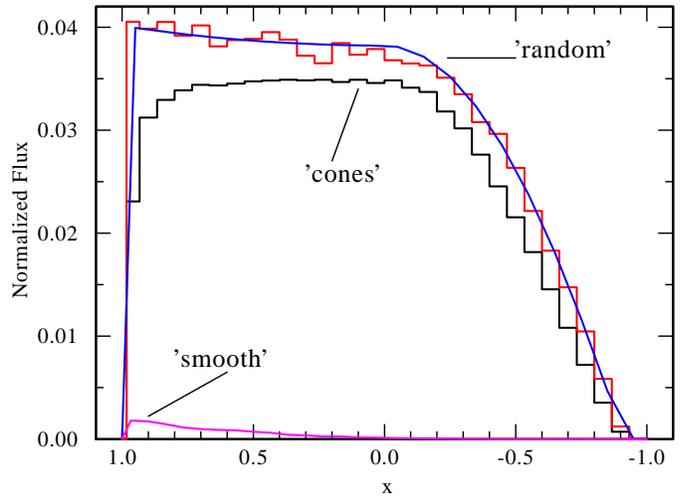}} 

\caption{Line profiles for a model with total optical depth $\tau_\ast
= 10$, and $N_r = 10$ shells in radial direction. The histogram-style
profiles are for the ``random fragments'' and the ``cones'' versions
of our stochastic wind model. The smooth line is analytically
calculated with the effective opacity from Eq.\,(\ref{eq:alpha}), and
is equivalent to the statistical model in Paper\,I. It agrees
precisely with the random fragment stochastic model. The more regular
arrangement of the absorbing fragments in the cones model leads to a
slightly lower emergent flux. In drastic contrast, a homogeneous
wind with the same mass-loss rate absorbs the X-ray line almost
totally (dashed line). Further model parameters are: constant
velocity; emission distributed between $r_{\rm em}^{\rm min} =
10\,R_\ast$ and $r_{\rm em}^{\rm max} = 30\,R_\ast$; absorbing shells
between $r_{\rm sh}^{\rm min} = 10\,R_\ast$ and $r_{\rm sh}^{\rm max}
= 100\,R_\ast$.}

\label{fig:n10}
\end{figure}

\subsection{Comparison with the analytical statistical model}
 
Solving the radiative transfer equation with the effective opacity
from Eq.\,(\ref{eq:alpha}) is equivalent to our statistical approach
in Paper\,I. In this subsection we discuss how this analytical
treatment compares with the stochastic model constructed in the
present paper.

\begin{figure}[hbtp] 
\centering 
\epsfxsize=\columnwidth
\mbox{\epsffile{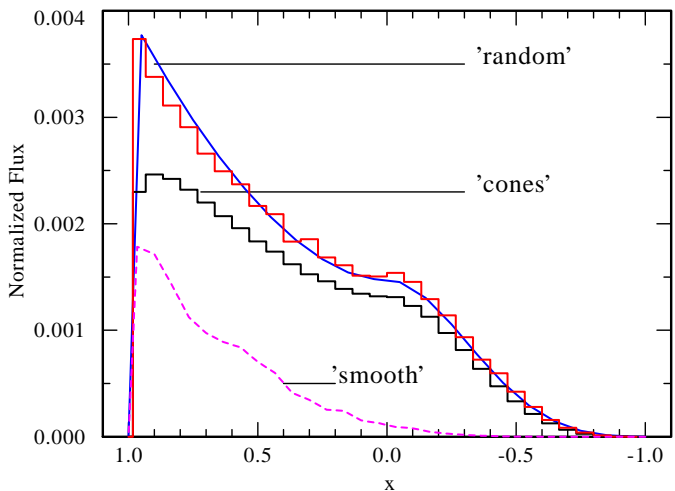}}
 
\caption{Same as Fig.\,\ref{fig:n10}, but with $N_r = 100$ fragments
in radial direction. Due to their larger number, most of the fragments
are now optically thin. Therefore the profiles from the stochastic
model are more similar to that from a homogeneous wind, although still
stronger by a factor of two.}

\label{fig:n100}
\end{figure}

Some resulting line profiles are shown in Figs.\,\ref{fig:n10} and
\ref{fig:n100}. The total optical depth is $\tau_\ast = 10$ for both
figures. The only difference is the number of fragments in radial
direction, which is $N_r = 10$ in Fig.\,\ref{fig:n10} and $N_r = 100$
in Fig.\,\ref{fig:n100}. The histogram-style profiles are for the
stochastic wind model in the ``random fragments'' and ``cones''
version. The smooth line is calculated analytically using the
effective opacity from Eq.\,(\ref{eq:alpha}), and is equivalent to
that from the statistical approach of Paper\,I. It is in excellent
agreement with the random fragment stochastic model, confirming its
prediction of a flat-topped, blueshifted and nearly symmetric
profile. The more regular arrangement of the absorbing fragments in
the cones model leads to a slightly lower emergent flux, especially at
the blue edge of the profile.

For the parameters in Fig.\,\ref{fig:n10}, most fragments are
optically thick. Therefore the gain in emergent flux achieved in the
stochastic wind model is dramatic. A homogeneous wind with the same
mass-loss rate absorbs the X-ray line almost totally (dashed line in
Fig.\,\ref{fig:n10}), and the tiny remaining line has a triangular
profile.

Figure\,\ref{fig:n100} shows the result for the same models, but with
the number of fragments in radial direction being increased to $N_r =
100$.  Now most of the fragments are no longer optically
thick. Therefore the emergent line profile from the stochastic wind
model (both versions) have a triangular shape like that from the
homogeneous wind, and the gain in emergent flux is only a factor
of two.

Thus our numerical results from the stochastic wind model
confirm fully the predictions from the statistical considerations in
Paper\,I. If the wind matter is compressed into a small number of
optically thick shells, and these shells break up into fragments, the
effective optical depth of the wind is small, independent of the
amount of mass contained in the fragments. On the contrary, a
homogeneous wind with the same mass-loss rate may have an arbitrarily
large optical depth. Therefore the gain in emergent intensity of an
X-ray line emitted from deep inside \ the wind may be huge, comparing
the fragmented with the homogeneous model.  The transparency of the
fragmented wind increases with decreasing number of fragments in
radial direction.

We emphasize that the line profiles from the fractured wind do {\em
not depend on the angular size of the shell fragments}. The number
$N_\vartheta$ of arcs in our 2-D cut of the model does not enter the
effective opacity. The statistical limit of a smooth line profile can
be achieved by a large $N_\vartheta$, but also by a large number of
random settings over which the results are averaged. Only for
technical reasons, the fragments should be small enough to justify the
neglect of curvature in our description. Moreover, in the cones model,
too big fragments in fixed angular positions would spoil the frequency
resolution and imprint spurious steps to the emergent profiles. 

\begin{figure}[hbtp] 
\centering 
\epsfxsize=\columnwidth
\mbox{\epsffile{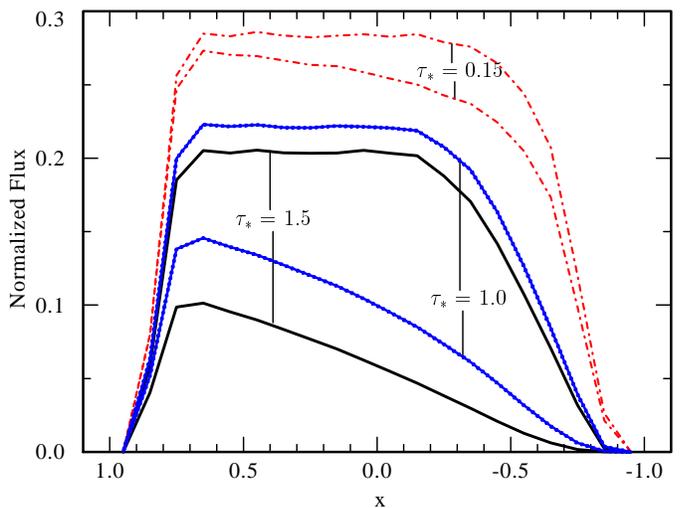}} 

\caption{Line profiles for relatively thin winds. The result from the
fragmented wind is compared to the corresponding homogeneous model
with the same mass-loss rate. The pairs are labeled with the total
optical depth of the homogeneous wind, $\tau_\ast$. The stronger line
of each pair refers to the fractured wind. The radial number of
fragments is always $\langle N_r \rangle = 1$. The differences between
the fractured and the homogeneous wind become significant when
$\tau_\ast \gtrsim 1$. The profiles from homogeneous models become
weaker and skewed, while the fragmented model profiles remain
flat-topped in the blue part. The velocity law exponent $\beta=1$.}

\label{fig:f1}
\end{figure}

\section{Parameter study}

In Sect.\,\ref{sect:basic-effects} we employed simplified models in
order to illustrate and understand the effects of wind fragmentation. In
this section line profiles are presented for realistic wind parameters,
calculated with our stochastic wind model in the ``random fragments''
version.

A stellar wind is specified by a few fundamental parameters, such as the
stellar luminosity, the photospheric radius, the mass-loss rate, the
chemical composition and the velocity law. Even for a homogeneous model,
the calculation of the continuous opacity at a given X-ray frequency is
not straightforward and requires in principle a Non-LTE atmospheric
model. Then one can calculate the radial optical depth $\tau_\ast$ of the
homogeneous model, as defined by Eq.\,(\ref{eq:tast}). As stated already
in Sect.\,\ref{sect:distribution_of_absorbers}, we assume for simplicity
that the X-ray opacities scale linearly with density, and adopt
$\tau_\ast$ of the corresponding homogeneous model as the free parameter
of our calculations.

The spatial structure of the inhomogeneous wind depends on the radial
distribution of the wind fragments and of the X-ray line emitting spots.
Important is the number of fragments in radial direction, $N_r$.
Existing hydrodynamical simulations trace the fragmentation to as far as
hundreds of stellar radii. X-ray emission is predicted to take place
only in the wind acceleration zone, i.e.\ out to a few stellar radii.

To keep the discussion transparent, we adopt the following
parameters for all figures in this section: a smooth wind between
$1\,R_\ast$ and $4\,R_\ast$, fragmentation of the wind between
$4\,R_\ast$ and $100\,R_\ast$ followed by a smooth wind again further
out; the X-ray line is emitted between $4\,R_\ast$ and $10\,R_\ast$. The
remaining parameters for the stochastic fragmented wind model are the
angular extension of the fragments and the number of fragments per
radius. The number of fragments in angular direction, for which we
generally use $N_\vartheta = 360$, is of no significance, as discussed
in the previous section.

Therefore, when $\tau_\ast$ is specified, the most influential parameter
of the model which is not constrained by independent means is the number
of fragments per radius, $\langle N_r \rangle$. This parameter defines
whether the fragments are optically thick or thin for the given
mass-loss rate. 

It was shown in Section\,5.2 that fragmentation becomes effective in
reducing opacity when fragments are optically thick, but cannot
influence the radiative transfer in optically thin winds even if the
number of fragments per radial direction, $N_r$, is small. To check
the parameter range we calculated some models with $\langle N_r
\rangle = 1$ and different values of $\tau_\ast$ around unity. The
profiles in Fig.\,\ref{fig:f1} confirm that the difference between the
homogeneous and fragmented wind is small for the optically thin case
$\tau_\ast = 0.15$, and only becomes significant when $\tau_\ast
\gtrsim 1$. With increasing $\tau_\ast$ the profiles from the
homogeneous models become weaker and skewed, while the fragmented
model profiles remain strong and flat-topped in the blue part.

\begin{figure}[hbtp] 
\centering 
\epsfxsize=\columnwidth
\mbox{\epsffile{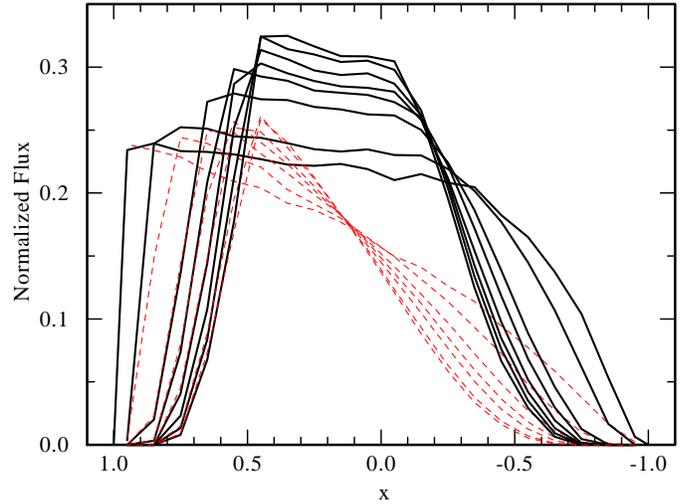}} 

\caption{Line profiles emerging from fragmented winds (solid lines)
and from smooth winds (dashed lines). All models are with $\tau_\ast =
1$ and $\langle N_r \rangle = 10$. The different profiles refer to
different velocity-law exponents $\beta =$ 0, 0.5, 0.8, 1, 1.5, 2,
2.5, 3, 3.5, and 4, where the higher $\beta$ values yield narrower
profiles.}

\label{fig:bvthin}
\end{figure}

\begin{figure}[hbtp] 
\centering 
\epsfxsize=\columnwidth
\mbox{\epsffile{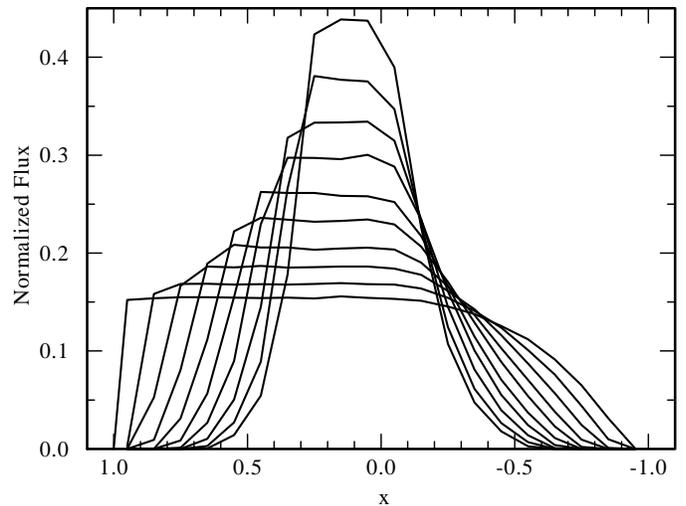}} 

\caption{Line profiles emerging from an optically thick wind, 
$\tau_\ast = 10$, compressed into one fragment per radial direction,
$\langle N_r \rangle = 1$. The line profiles are for different velocity 
laws, namely for constant velocity, and exponents $\beta =$ 0.5, 1, 1.5, 2, 
2.5, 3, 3.5, 4, 4.5 (from lower to higher maximum).  
}
\label{fig:bvthick}
\end{figure}

The next two figures study the influence of different velocity laws on
the line profile. First we consider a relatively weak wind, with
optical depth $\tau_\ast = 1$. The fragmented wind is calculated with
$\langle N_r \rangle = 10 $ shells in radial direction.  The line
profiles shown in Fig.\,\ref{fig:bvthin} are for different values of
$\beta = $ 0.5, 0.8, 1, 1.5, 2, 2.5, 3, 3.5 and 4, respectively. The
broadest profile is for the constant velocity case, which can be
considered as the limit $\beta \rightarrow 0$ when $v(r)$ jumps to
$v_\infty$ immediately at the stellar surface. The higher $\beta$, the
slower is the wind acceleration.  CAK-type hydrodynamic modeling of
O-star winds arrives at $\beta \approx 0.8$, typically, while certain
empirical facts have been interpreted in favor of higher $\beta$
values. The X-ray line profiles from our stochastic wind model show
some sensitivity to this parameter.

At the opposite extreme, Fig.\,\ref{fig:bvthick} shows an optically
thick wind ($\tau_\ast = 10$) compressed into just one fragment per
radial direction $\langle N_r \rangle = 1$. The line profiles are
again for different velocity-law exponents $\beta$ ranging from
constant velocity ($\beta \rightarrow 0$) to $\beta = 4.5$. For high
values of $\beta$ the lines are nearly symmetric, and only slightly
shifted. Due to the high optical depth, the emergent lines from the
corresponding homogeneous models are far too weak to be plotted to the
same scale.

\begin{figure}[hbtp] 
\centering 
\epsfxsize=\columnwidth
\mbox{\epsffile{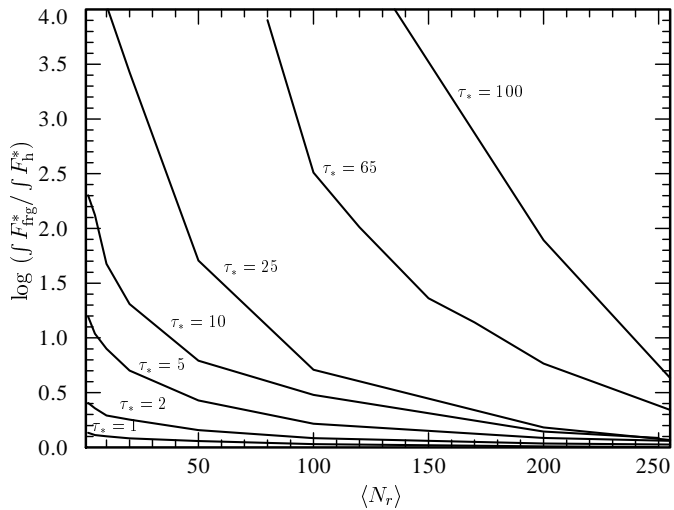}} 

\caption{The gain in emergent line flux by wind fragmentation. The
ratio of the integrated line flux from fragmented and homogeneous wind
models, ${\cal R}=F^\ast_{\rm frg}/F^\ast_{\rm h}$, is shown on a
logarithmic scale. The curves show model series with different optical
depths $\tau_\ast$ (labels).  Along each series the number of shells
in radial direction, $\langle N_r \rangle$, is varied. The gain factor
${\cal R}$ can be huge for very thick winds, i.e.\ high $\tau_\ast$,
but is reduced with increasing number of radial fragments, $\langle
N_r \rangle$. All models are for a velocity-law exponent $\beta=0.8$.}

\label{fig:ratio}
\end{figure}

The most important consequence of the wind fractioning is the  drastic
reduction of the wind opacity. Therefore, emergent  X-ray
line fluxes can be much higher than from homogeneous winds of the same
mass-loss rate. The gain factor, i.e.\  the integrated emergent line
flux ratio between a fragmented wind and the corresponding  homogeneous
wind, ${\cal R}=\log\,F^\ast_{\rm frg}/F^\ast_{\rm h}$, is shown in
Fig.\,\ref{fig:ratio}. The fluxes are integrated over the whole line
profile. The figure displays the dependence of ${\cal R}$ on the
radial number of fragments, $\langle N_r \rangle$. The different series
of models have different optical depths $\tau_\ast$ (labels). The plot
demonstrates that fragmentation can increase the emergent line flux by
orders of magnitude. The effect is extremely pronounced for very thick
winds, i.e.\ high $\tau_\ast$. Radial fragmentation into very many
shells reduces the gain factor.

\section{Discussion and conclusions}

\begin{figure}[bth] 
\centering 
\epsfxsize=\columnwidth
\mbox{\epsffile{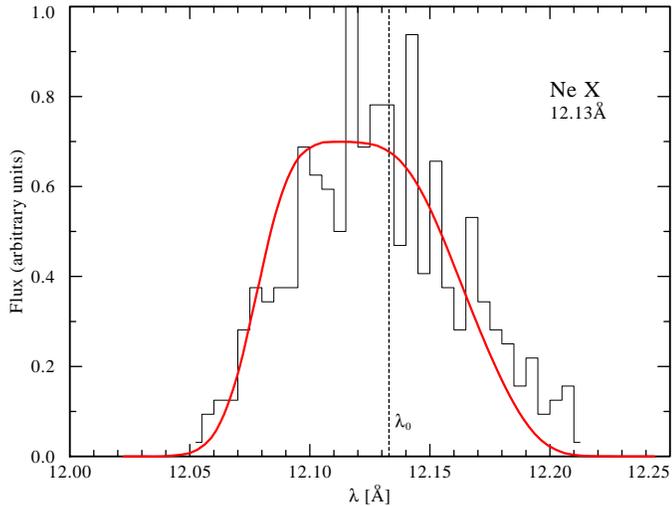}} 

\caption{Profile of the Ne\,{\sc x} line from $\zeta$\,Orionis, as
observed with the Chandra satellite (histogram). The smooth line is a
tentative comparison with one of our calculations, taking the profile
for $\tau_\ast = 10$ and $\beta = 2$ from Fig.\,\ref{fig:bvthick}. The
theoretical profile is scaled to the terminal wind velocity of $\zeta$
Orionis (2100\,km\,s$^{-1}$) and convolved with the instrumental
profile, a Gaussian of 0.023\,\AA\ FWHM. The good fit demonstrates
that our fragmented wind model can reproduce emergent line profiles of
the observed shape.}

\label{fig:zeta_Ori}
\end{figure}

A main motivation for our fragmented wind model is to explain the
observed X-ray line profiles from O stars, which cannot be reproduced
so far. We may check now whether we came closer to this goal.
Figure\,\ref{fig:zeta_Ori} shows the profile of the Ne\,{\sc x} line
from $\zeta$\,Orionis, as observed with the Chandra satellite
(histogram). The smooth line is a tentative comparison with one of our
calculations. From the known mass-loss rate of the star, the continuum
at the frequency of this line may be optically thick (Waldron et al.\
\cite{WaldCass01}).  Therefore we select a suitable profile (with
$\beta = 2$) from the set in Fig.\,\ref{fig:bvthick} with $\tau_\ast =
10$. The dimensionless frequency is converted into wavelength
according to the terminal wind velocity of $\zeta$\,Orionis
(2100\,km\,s$^{-1}$), and the profile is convolved with a Gaussian of
0.023\,\AA\ FWHM to account for instrumental broadening. The good
agreement just demonstrates that our fragmented wind model can
reproduce emergent line profiles of the observed shape, even if
$\tau_\ast$ is large. For a systematic fit, all available line
profiles should be considered in order to adjust the parameters of the
model consistently. This will be subject of a forthcoming paper.

There are two aspects in the problem of fitting the observed line
profiles. One is to reproduce the line shape, another is to provide
for the observed line flux. Ignace \& Gayley (\cite{Rico02})
considered profile shapes for optically thick emission lines. They
showed that these lines are blueshifted, and slightly narrower than
optically thin lines. The lines have a universal shape which is nearly
symmetric and insensitive to the level of continuous
absorption. Nevertheless, the flux in the line scales inversely with
optical depth of the cool wind, and therefore the problem of missing
opacity is still unresolved in their approach. The same problem is
highlighted in Kramer et al.\ (\cite{KCO03}), which up-to-day is the
most consistent attempt to fit the observed lines. Our study shows
that incorporation of wind fragmentation can resolve the controversy.

A burning question is, why stars with existing high-resolution X-ray
spectra show qualitatively different profile shapes, as listed in the
introduction. At present we suggest that only $\zeta$\,Pup is a safe
prototype for an isolated strong stellar wind.  Indeed, speckle
interferometry led to the discovery that $\theta^1$\,Ori\,C and
$\zeta$\,Ori\,A, are binary systems (Hummel et al.\ \cite{Hummel00},
Weigelt et al.\ \cite{Weigelt99}), and $\delta$\,Ori\,A is a
well-known multiple system (see e.g. Miller et al.\
\cite{Milleretal02}). Moreover, $\theta^1$\,Ori\,C has a strong
magnetic field in excess of 1\,kG (Donati et al.\
\cite{Don02}). The physical processes leading to the emission of
X-rays can differ between objects. Nevertheless, we claim that the
effects of wind fragmentation ought to be included in the modeling of
emerging line profiles independent of the particular emission
mechanism.

The scenario chosen as a framework of the model presented here is that
of Feldmeier et al.\ \cite{AF97}, where X-ray are produced in the
collision of fast cloudlets with dense shells. The prominent feature
of this hydrodynamic simulation is that the X-ray emitting plasma is
always located at the starward face of the cool absorbing
fragments. It was already shown in Paper\,I, and is confirmed here,
that this configuration leads to strong depletion of the central part
of the line, an effect which is certainly not observed.

The following conclusions can be drawn  from our numerical
modeling of the X-ray line emission from inhomogeneous, fragmented
stellar wind:

\begin{enumerate}

\item Fragmentation drastically reduces the effective opacity of the
wind. Therefore X-rays produced deep inside the wind can effectively
escape, which would be totally absorbed in a homogeneous wind of the
same mass-loss rate.

\item Absorption in a fragmented wind becomes effectively independent
of the mass absorption coefficient and therefore of the wavelength, if
the fragments are optically thick.

\item The line profiles from the fragmented wind model exhibit a
variety of shapes, ranging from broad, blueshifted and flat-topped to
narrow and nearly symmetric, and are thus of promising similarity to
observations.

\item The possibility of a flat-topped, i.e.\ almost symmetric blue
part of the line profile is due to the model assumption that the
fragments are flat, have small thickness, and are aligned
perpendicular to the radial flow direction.

\item The effect of fragmentation is significant when the individual fragments
are optically thick. Therefore the effect depends on the average number
of fragments per radial direction. This parameter can be empirically
restricted by a detailed line fit and has interesting implications for
the theoretical understanding of the wind hydrodynamics.

\item
The lateral size of the wind fragments has no influence on the emergent 
lines. 

\item 
If the fragments are strictly confined to a pattern of radial cones
while moving outwards, which ensures strict mass conservation in each
radial direction, the resulting line profiles differ only little from
the random fragment model where the absorbing shell fragments are
randomly distributed in both radial and angular coordinates.

\item The numerical modeling featuring stochastically arranged
emitting parcels of gas and absorbing fragments confirms the results
of Paper\,I, where analytical formulae have been derived from a
statistical treatment.

\end{enumerate}

\begin{acknowledgements}
L.M.O.\ acknowledges support from a Deutsche Forschungsgemeinschaft 
grant (Fe 573/1-1).
\end{acknowledgements}


\end{document}